\documentclass[aps,prl,reprint,superscriptaddress,floatfix]{revtex4-1}
\usepackage{graphicx} % For figures
\usepackage{color,dcolumn,bm,amssymb,amsmath} % Math and symbols
\usepackage{soul,xcolor} % Text formatting
\usepackage{siunitx} % SI units
\usepackage{epstopdf,gensymb} % Graphics support
\usepackage{hyperref} % For hyperlinks in references
\DeclareSIUnit\angstrom{\text{\AA}} % Angstrom unit
\usepackage{siunitx} % SI units
\begin{document}

% Title
\title{Imaging the Meissner effect and local superfluid 
stiffness in a graphene superconductor}

\author{Ruoxi Zhang}
\thanks{These authors contributed equally.}
\affiliation{Department of Physics, University of California, Santa Barbara, California 93106, USA}

\author{Benjamin A. Foutty}
\thanks{These authors contributed equally.}
\affiliation{Department of Physics, University of California, Santa Barbara, California 93106, USA}
\author{Owen Sheekey}
\thanks{These authors contributed equally.}
\affiliation{Department of Physics, University of California, Santa Barbara, California 93106, USA}
\author{Trevor Arp}
\affiliation{Department of Physics, University of California, Santa Barbara, California 93106, USA}
\author{Siyuan Xu}
\affiliation{Department of Physics, University of California, Santa Barbara, California 93106, USA}
\author{Tian Xie}
\affiliation{Department of Physics, University of California, Santa Barbara, California 93106, USA}
\author{Yi Guo}
\affiliation{Department of Physics, University of California, Santa Barbara, California 93106, USA}
\author{Hari Stoyanov}
\affiliation{Department of Physics, University of California, Santa Barbara, California 93106, USA}
\author{Sherlock Gu}
\affiliation{Department of Physics, University of California, Santa Barbara, California 93106, USA}
\author{Aidan Keough}
\affiliation{Department of Physics, University of California, Santa Barbara, California 93106, USA}
\author{Evgeny Redekop}
\affiliation{Department of Physics, University of California, Santa Barbara, California 93106, USA}
\author{Canxun Zhang}
\affiliation{Department of Physics, University of California, Santa Barbara, California 93106, USA}
\author{Takashi Taniguchi}
\affiliation{Research Center for Materials Nanoarchitectonics, National Institute for Materials Science, 1-1 Namiki, Tsukuba 305-0044, Japan}
\author{Kenji Watanabe}
\affiliation{Research Center for Electronic and Optical Materials, National Institute for Materials Science, 1-1 Namiki, Tsukuba 305-0044, Japan}
\author{Martin E. Huber}
\affiliation{Departments of Physics and Electrical Engineering, University of Colorado, Denver, Colorado 80204, USA}
\author{Chenhao Jin}
\affiliation{Department of Physics, University of California, Santa Barbara, California 93106, USA}
\author{Erez Berg}
\affiliation{Department of Condensed Matter Physics, Weizmann Institute of Science, Rehovot 76100, Israel}
\affiliation{Materials Department, University of California Santa Barbara, Santa Barbara 93106 USA}
\affiliation{Department of Electrical and Computer Engineering, University of California, Santa Barbara, CA 93106,
USA}
\author{Andrea F. Young}
\email{andrea@physics.ucsb.edu}
\affiliation{Department of Physics, University of California, Santa Barbara, California 93106, USA}

\begin{abstract}
We report the observation of the Meissner effect in a rhombohedral graphene superconductor, realized via direct imaging of the static fringe magnetic field. 
In our few-micron sample, the onset of superconductivity manifests as a diamagnetic response that screens only $\sim 100$ ppm of the applied magnetic field. 
Tracking the evolution of the resulting nanotesla-scale fields in real space allows us to observe the entry of superconducting vortices and map the local superfluid stiffness, $\rho_s$.   
Correlating fringe field signals from both Meissner screening and magnetically ordered states, we show that superconductivity onsets in the midst of a continuous quantum phase transition to a canted spin ferromagnet.
Within the superconducting state, we find the temperature dependence of $\rho_s$ to be incompatible with isotropic Bardeen-Cooper-Schrieffer theory and the zero-temperature stiffness $\rho_s^0$ to be linearly proportional to $T_c$, constraining future theoretical models of superconductivity in this system.
\end{abstract}

\maketitle

\section{Introduction}
The expulsion of magnetic flux, known as the Meissner effect, is the key thermodynamic property that distinguishes the superconducting state from that of a ballistic metal.  
While the detection of Meissner effects is part of the standard toolkit for characterizing three--dimensional (3D) superconductors, for two--dimensional (2D) superconductors the magnitude of the screening current is strongly suppressed. 
Indeed, in low density 2D superconductors such as moire and rhombohedral graphene, Meissner effects have yet to be directly observed. 

In a 2D superconductor, the diamagnetic response---captured by the screening current density $\mathbf{j}_s$---can be understood within the London limit of Ginzburg-Landau theory as
\begin{equation}
\label{eq:GL_London}
\mathbf{j}_s = \frac{2\pi k_B\rho_s}{\Phi_0} \left(\nabla\phi-\frac{2\pi}{\Phi_0}\mathbf{A}\right),
\end{equation}
where $k_B$ is the Boltzmann constant, $\phi$ is the superconducting phase, $\mathbf{A}$ is the electromagnetic vector potential, and $\rho_s$ is the superfluid stiffness, which parameterizes the energy cost of phase deformations (here $\Phi_0\approx\qty{2e-3}{T\cdot nm^2}$ is the superconducting flux quantum) \cite{tinkham_introduction_2015}.
In addition to $\rho_s$, the strength of the Meissner effect is controlled by the sample geometry.  As a concrete example,  immediately above a small disc of finite radius $a$ in the presence of an applied field  $B_z$, the screening currents induce a change in the fringe magnetic field $\Delta B$ given by 
\begin{equation}
\Delta B/B_{z} \approx -a/(2\Lambda),
\end{equation}
where $\Lambda=\frac{\Phi_0^2}{2\pi^2\mu_0 k_B }\frac{1}{\rho_s}\approx \frac{1.2\rm{cm\cdot K}}{\rho_s}$ is known as the Pearl length \cite{pearl_current_1964,fetter_flux_1980}, and $\mu_0$ is the magnetic permittivity of free space.   
This is a far weaker effect than obtained in a 3D superconductor, where Meissner screening becomes perfect in the bulk over a much smaller length scale, known as the London penetration depth $\lambda_L$. $\lambda_L$ can be recovered from the 2D limit by considering a semi-infinite stack of 2D sheets with Pearl length $\Lambda$ and interlayer spacing $d$ as $\lambda_L^2= \Lambda d/2$. For layered superconductors with interlayer separation on the angstrom scale, then, $\lambda_L$ may be smaller by several orders of magnitude, making the Meissner effect far more readily observable.  

\begin{figure*}[ht!]
\centering
\includegraphics[width=\textwidth]{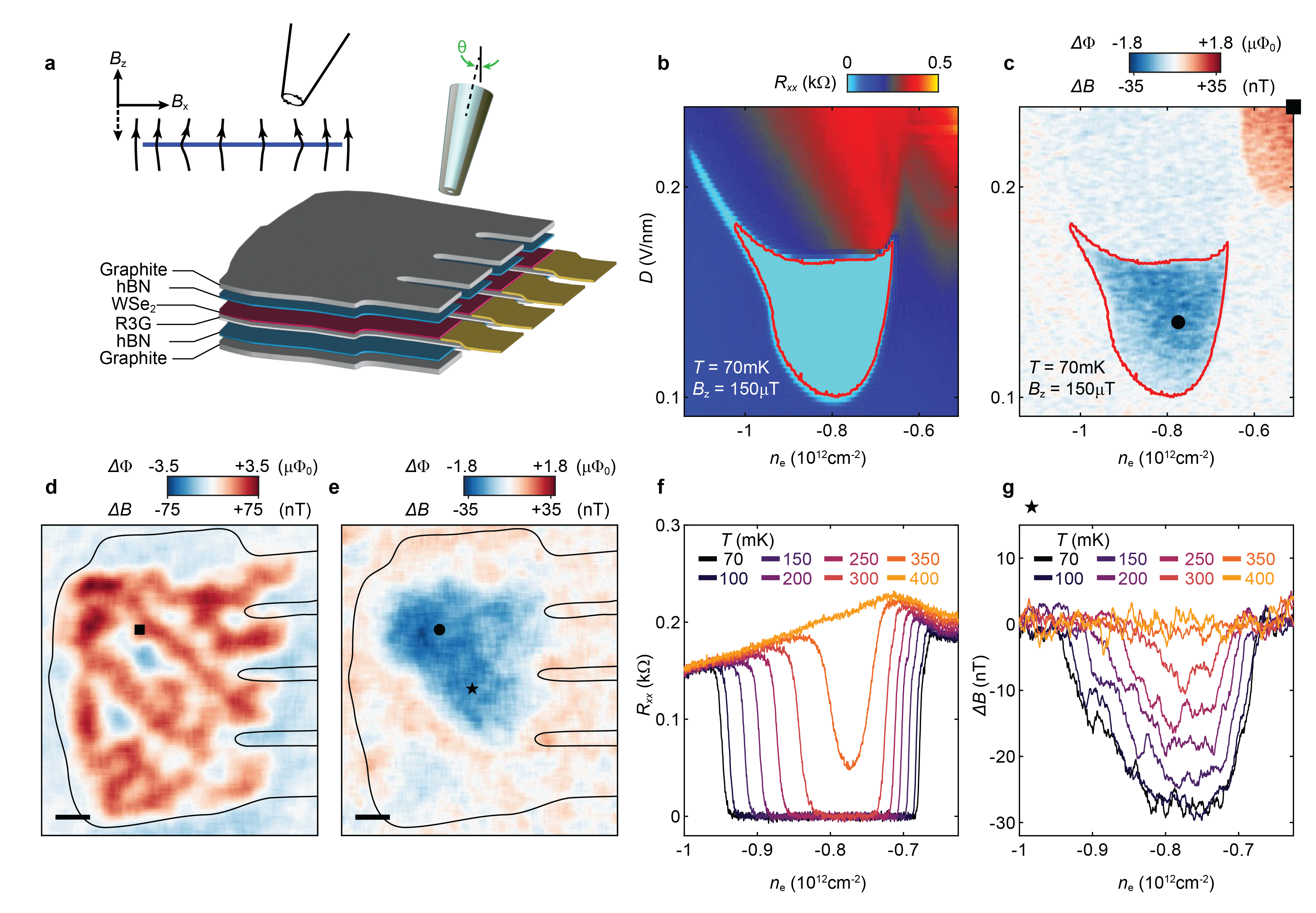}
\caption{\textbf{Meissner effect in R3G/WSe$_2$.}
\textbf{a}, Measurement and device schematics. The nSOT is mounted at an angle $\theta$ relative to the $z$--axis, which is perpendicular to the sample plane. 
Transport experiments are realized with four contacts arranged on the right side of the device. 
The inset illustrates the applied external magnetic fields in the x- and z-directions as well as the fringe field pattern due to the Meissner effect.
\textbf{b}, $R_{xx}$ measured as a function of $n_e$ and $D$. The light cyan region corresponds to the superconducting state, characterized by vanishing resistance. 
The red contour corresponds to $R_{xx}=5\Omega$.
\textbf{c}, Fringe magnetic field $\Delta B$ measured at a fixed location above the sample
A diamagnetic Meissner signal (rendered in blue) is associated with the superconducting region of panel b. A ferromagnetic region (rendered in red) is also visible at the upper right. 
\textbf{d}, Spatial image of $\Delta B$ within the ferromagnetic state corresponding to the $n_e, D$ configuration marked by the solid square in panel c. 
An outline of the device is shown in overlay as a guide to the eye, while the solid square indicates the spatial position where data in panel c were acquired. The scale bar is $1\mu$m throughout this work. 
\textbf{e}, Spatial image of $\Delta B$ in the superconducting state, corresponding to the $n_e, D$ configuration marked by the solid circle in panel c. 
\textbf{f}, $R_{xx}$ at fixed $D=0.138V/nm$ as a function of $n_e$ for different values of $T$. 
\textbf{g}, $\Delta B$, measured at the spatial point marked by the solid star in panel e, for the same parameters as the data in panel f. 
}
\label{fig:fig1}
\end{figure*}

Despite this suppression, diamagnetic response has been successfully observed in a number of 2D superconductors.  
In a pioneering experiment, Tafuri and Kirtley used a scanning superconducting quantum interference device (SQUID) to measure $\Lambda$ from the spatial structure of magnetic screening in the vicinity of a `Pearl vortex' in a cuprate thin film \cite{tafuri_magnetic_2004}.
Subsequent experiments have used scanning SQUID microscopy \cite{finkler_scanning_2012, persky_studying_2022} or scanning nitrogen-vacancy magnetometry \cite{rondin_magnetometry_2014, casola_probing_2018} to image the Meissner effect in 2D superconductors \cite{bert_direct_2011,bert_gate-tuned_2012,thiel_quantitative_2016,jarjour_superfluid_2023,fridman_anomalous_2025}.
Notably, these experiments were aided by comparatively large superfluid stiffness and sample sizes, enabling relatively large magnetic field signals. 

Flat band graphene superconductors have attracted attention due to the proximity between superconductivity and correlated magnetic phases as well as their electric-field effect tunability \cite{cao_unconventional_2018,yankowitz_tuning_2019,park_tunable_2021,hao_electric_2021,zhang_promotion_2022,park_robust_2022,zhou_superconductivity_2021,zhou_isospin_2022,han_signatures_2025}.  
Many outstanding questions motivate the development of thermodynamic probes of the superconducting state.
In most graphene superconductors, the structure of the superconducting gap and the underlying pairing mechanism remain unknown; both questions could be addressed by measurements of $\rho_s$ via the Meissner effect.   
However, measuring the Meissner effect in graphene systems is experimentally challenging due to the energetics of the flat bands.  
In two dimensions, $\rho_s$ is bounded from below by its value at the Berezinskii--Kosterlitz--Thouless (BKT) transition where $\rho_s(T_c)= \frac{2}{\pi} k_B T_c$ \cite{berezinskii_destruction_1970,kosterlitz_ordering_1973}.
In Bardeen-Cooper-Schrieffer (BCS) framework, the $T\rightarrow 0$ limit of $\rho_s$ is set by the Fermi energy $E_F/(8\pi)=\hbar^2 n/ (8 m^*)$, where $n$ is the carrier density and $m^*$ is the effective mass, limiting the strength of the Meissner response in flat band systems where the expected $E_F\lesssim 10 \rm{ meV}$.
Compounding the experimental difficulty is the small size of typical samples, which further limits the screening field strength.  
An alternate experimental approach used in two recent experiments \cite{banerjee_superfluid_2025,tanaka_superfluid_2025} leveraged the high frequency complex conductivity to infer $\rho_s$ in moire graphene. 
However, the micron scale inhomogeneity typical of low-density electron systems complicates the interpretation of the kinetic inductance in terms of intrinsic properties of the superconducting state observed in transport. 
Direct, local measurements of $\rho_s$ via the Meissner effect thus remain a highly desirable experimental goal. 

\section{Meissner effect in R3G/WSe\textsubscript{2}}

Here, we use scanning nanoSQUID-on-tip (nSOT) microscopy to locally probe Meissner screening in a rhombohedral graphene superconductor. Our experimental geometry is shown in Fig. \ref{fig:fig1}a. The sample consists of a dual-graphite gated rhombohedral graphene trilayer (R3G) flake clad on one side with a WSe\textsubscript{2} monolayer and connected to four transport contacts. To map the fringe magnetic fields, we raster an ultra-sensitive nSOT tip with $1-3$ $\rm{nT/\sqrt{Hz}}$ sensitivity (see Extended Data Fig. \ref{fig:nSOTChar}) in a fixed-height plane above the sample. The experiment is mounted in a hermetic cell bolted to the mixing chamber of a dilution refrigerator with a nominal base temperature of 30mK.  In order to probe the superconducting phase diagram near $B_z=0$, we fix the nSOT tip at an angle of $12.5^\circ$ relative to the axis normal to the sample plane, and flux-bias the nSOT to its sensitive point with a constant $46$ mT magnetic field applied in the plane of the sample.  The perpendicular applied field $B_z$ can then be tuned over a continuous range from $B_z=-250\ \rm{\mu T}$ to $B_z=750\ \rm{\mu T}$.  
To extract the magnetic signal, we apply a square wave to the bottom gate at a frequency $f=$ 247Hz or 281 Hz, modulating the sample between the $n_e$ and $D$ point to be measured and a confirmed nonmagnetic phase.  Demodulating the resulting nSOT signal isolates the field-effect tuned contributions to the magnetic field, giving the desired static screening magnetic field, $\Delta B(x,y,n_e,D)$ (see Methods and Extended Data Fig. \ref{fig:Boxcar}).

Fig. \ref{fig:fig1}b shows $R_{xx}$ measured at $T=70$ mK and $B_z=150\mu T$.  Consistent with prior work\cite{patterson_superconductivity_2025,yang_impact_2025}, a robust zero resistance state is observed over a finite range of $D$ and $n_e$ when carriers at the Fermi level are polarized to the layer closest to the WSe$_2$ (see Extended Data Fig. \ref{fig:PhaseDiagrams}).  Fig. \ref{fig:fig1}c shows the measured $\Delta B$ at a fixed position over the sample as a function of $n_e$ and $D$.  We present the data in both magnetic field units as well as in units of the magnetic flux threading our 350 nm diameter nSOT, $\Delta \Phi$.  
The superconducting region is characterized by a negative $\Delta B$--i.e., we observe a reduction in the magnetic field relative to the applied $B_z=150\mu T$. 
In addition, a distinct region of positive $\Delta B$ is observed at upper right; this region is associated with an orbital ferromagnetic `valley imbalanced' state (see Extended Data Fig. \ref{fig:PhaseDiagrams} and Ref. \cite{patterson_superconductivity_2025}).  To confirm the contrasting nature of the superconducting and ferromagnetic responses, Fig. \ref{fig:fig1}d shows a real space map of $\Delta B$ in  the valley imbalanced phase. Most of the sample area shows fringe field enhancement, consistent with orbital magnetic moments polarized in the direction of the applied $B_z$.  In contrast, in the superconducting state, $\Delta B$ is negative over a several-micron wide spatial domain, consistent with the out-of-plane diamagnetic response expected for Meissner screening. The maximal diamagnetic screening field observed in Fig. \ref{fig:fig1}e is approximately 30 nT, 5,000 times smaller than the applied $B_z$.  Given the approximate superconducting domain radius of 2 $\rm{\mu m}$, we infer a Pearl length of $\Lambda \approx 5 \rm{mm}$. Surprisingly, the spatial pattern of disorder in the valley imbalanced phase is distinct from what we observe in the superconducting state. 
As an additional confirmation that the magnetic response arises from the superconducting state, Figs. \ref{fig:fig1}f-g compare the temperature dependence of $R_{xx}$ and $\Delta B$ measured above the diamagnetic patch.  The diamagnetic signal onsets with the zero resistance state as a function of both $n_e$ and $T$.  

\begin{figure*}[ht!]
\centering
\includegraphics[width=18cm]{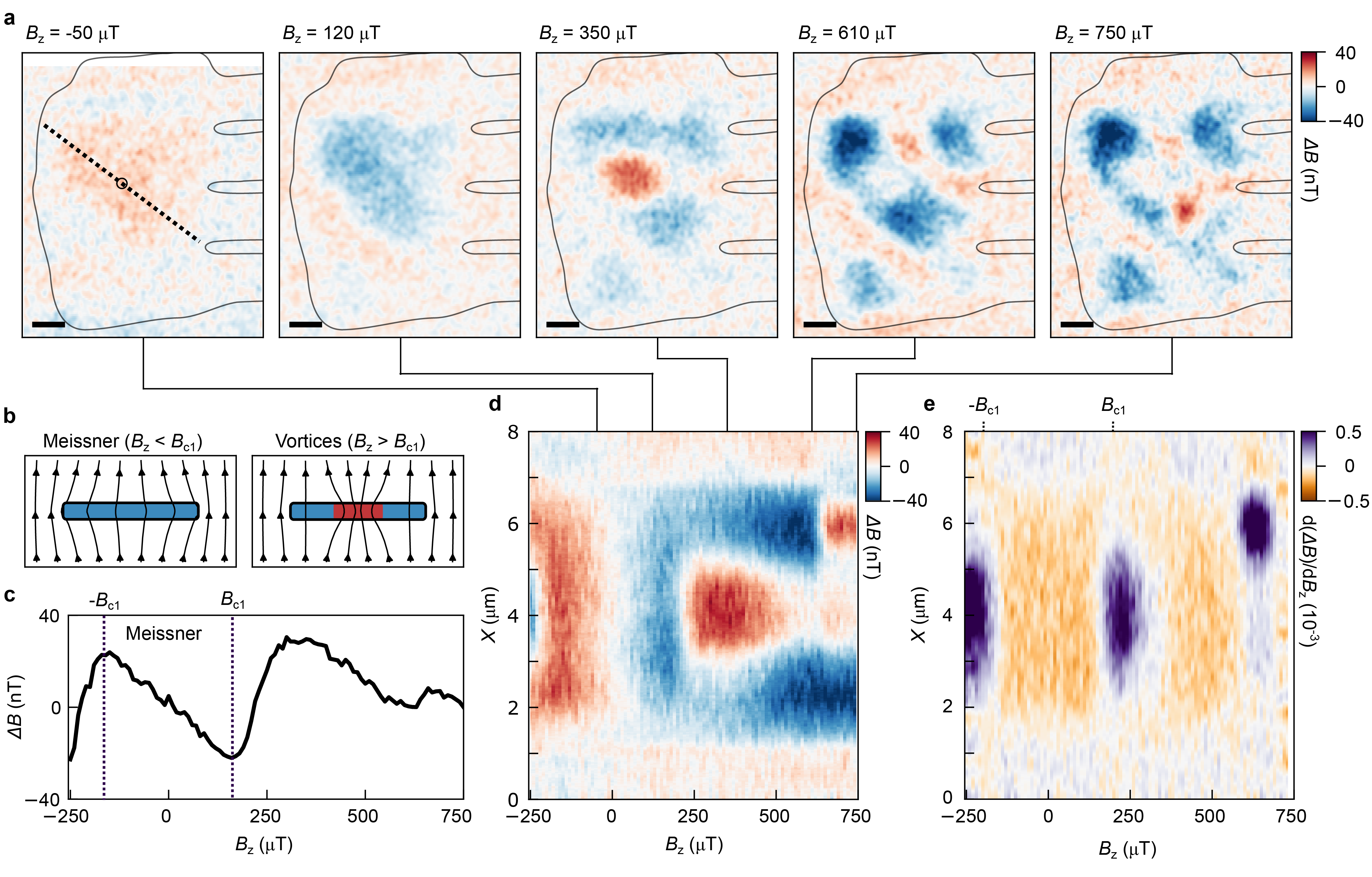}
\caption{\textbf{Spatial imaging of vortices and direct measurement of $B_{c1}$.}
\textbf{a}, Spatial images of $\Delta B$ at different values of applied $B_z$. Scale bar is 1 $\rm{\mu}$m. 
$n_e$ and $D$ are chosen at the center of the superconducting region (solid circle in Fig. \ref{fig:fig1}c). 
\textbf{b}, Schematic of magnetic field focusing due to screening currents in the pure Meissner state ($B_z<B_{c1}$, left) and in the presence of a vortex ($B_z>B_{c1}$, right).  
\textbf{c}, $\Delta B$ as a function of $B_z$ measured at the spatial position indicated by the circle in panel a. 
Near $B_z = 0$, the behavior is linear consistent with a constant diamagnetic susceptibility.  At $B_z = \pm B_{c1} \approx 200$ $\rm{\mu} T$, the flux changes sharply due to the appearance of a localized vortex.
\textbf{d}, $\Delta B$ measured along the dotted line in panel a, for variable $B_z$. $X$ denotes the coordinate along the dotted line with $X = 0$ in the upper left of the image in panel a.  
\textbf{e}, Numerical derivative $\rm{d}(\Delta B)/\rm{d} B_z$ for the data in panel d.  Orange regions denote constant negative susceptibility, while purple regions are associated with the entry of a new localized vortex to the sample.  
}
\label{fig:fig2}
\end{figure*}

\section{Vortices and lower critical field}

The energetics of Meissner screening also control the formation of superconducting vortices. Within Ginzburg--Landau (GL) theory, the balance between magnetic-field expulsion and the cost of creating normal--superconducting interfaces controls how the Meissner response evolves with increasing magnetic field. Due to their small $\rho_s$, 2D superconductors are strongly in the ``type-II" limit, lowering their free energy by admitting flux as quantized vortices above a lower critical field $B_{c1}$. 
The onset and spatial organization of these vortices can provide a sensitive local probe of sample inhomogeneity. For example, preferential pinning sites reveal minima in the vortex free-energy landscape. While electronic transport measurements have shown evidence for vortices in graphene superconductors\cite{perego_experimental_2025,perego_pearl-vortex_2026}, direct detection has been difficult because $\rho_s$ sets the scale of the magnetic vortex signal \cite{tafuri_magnetic_2004,fridman_anomalous_2025}.

With this motivation, we study the evolution of the Meissner fringe field with applied $B_z$ (Fig. \ref{fig:fig2}a, Extended Data Fig. \ref{fig:all_b_dep}). For small fields, $|B_z| < 200$ $\mu$T, we confirm that the magnetic signal reverses sign around $B=0$ and remains approximately uniform in shape. As $|B_z|$ increases further, the spatial pattern of the magnetic response becomes more complex. While some areas of the device have a growing diamagnetic response (darker blue in Fig. \ref{fig:fig2}a), in others $\Delta B$ reverses sign. We associate the sign-reversal with superconducting vortices, as the applied magnetic flux is focused through the normal core of the vortex (Fig. \ref{fig:fig2}b). 

This picture is made clearer by tuning $B_z$ continuously. In Fig. \ref{fig:fig2}c, we show $\Delta B$ measured at a single spatial position (open circle in Fig. \ref{fig:fig2}a) as a function of $B_z$. Around $B_z=0$, the dependence is linear with a vanishing signal at $B_z=0$, confirming the negative susceptibility of Meissner diamagnetism. 
At $|B_z| \approx 200$ $\rm{\mu}$T, we see a sharp change in $\Delta B$, localized over a small range of $B_z$ and associated with the appearance of a positive vortex signal in Fig. \ref{fig:fig2}a. We identify this value of $B_z$ as the empirically determined lower critical field $B_{c1}$ where the first vortex enters the sample. 
In Fig. \ref{fig:fig2}d-e, we plot $\Delta B$ and its numerical derivative $\rm{d}(\Delta B)/\rm{d}B_z$ for varying real space coordinate corresponding to the dotted line in Fig. \ref{fig:fig2}a. The numerical derivative clearly disambiguates the linear-in-$B_z$ diamagnetic Meissner response (rendered in orange) from the sudden entry of localized vortices (rendered in purple).  Notably, similar Meissner screening is evident at the same spatial position as the 1st vortex even above $B_{c1}$.  
This is as expected for the limit in which $\Lambda$ far exceeds the characteristic length scales in the sample: it is well-established both theoretically \cite{bardeen_quantization_1961,fetter_flux_1980,kogan_pearls_1994} and experimentally \cite{geim_non-quantized_2000,zhang_tunable_2022} that vortices in this limit have non-quantized magnetic flux, though the fluxoid (including current contributions) remains quantized.

Because of the finite-frequency modulation of the gates, the superconducting phase is destroyed and reformed at fixed applied $B_z$ many times for each measured data point. 
Our results thus correspond to an average over many field-cooling cycles.  
The observed pattern, with clearly localized vortex positions, implies that the pinning sites for the first few vortices are not random, but rather reflect consistent minima in the vortex free energy landscape. One evident trend is that the pinning sites are correlated with spatial positions where the fringe field signal of the nearby valley imbalanced phase is negative, as can be seen in Extended Data Fig. \ref{fig:PhaseDiagrams}a-d, pointing to a correlation between vortex pinning and mesoscopic inhomogeneity in the sample. It is notable that this inhomogeneity does not obviously manifest in the Meissner response near $B=0$; evidently, the zero vortex superconducting state, vortex pinning sites, and valley polarized magnetic states respond differently to spatially varying moire potential, spin orbit coupling, or other sources of micron-scale inhomogeneity in the underlying sample structure. 

\begin{figure}[ht!]
\centering
\includegraphics[width=9cm]{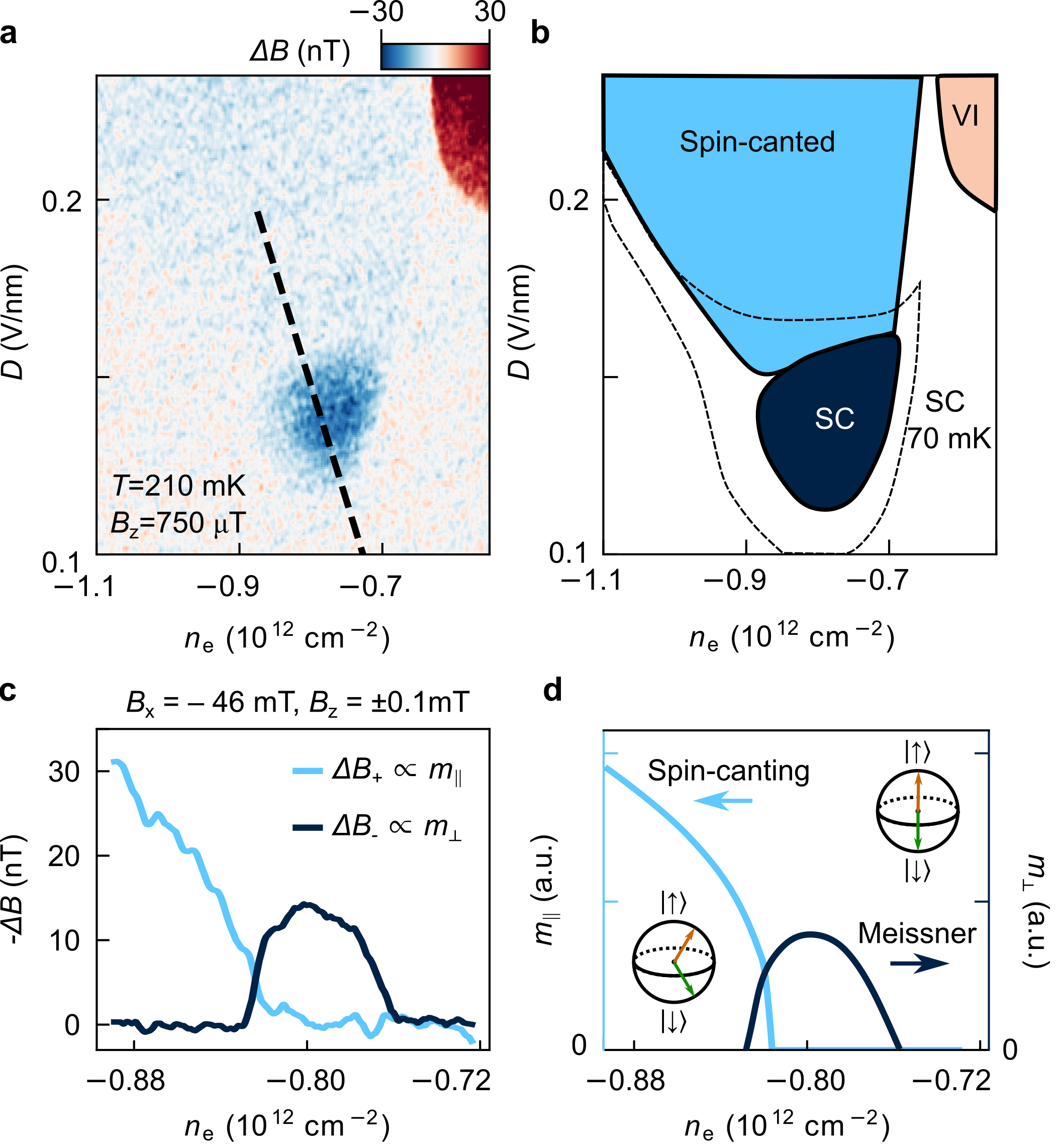}
\caption{\textbf{Relationship of spin-canted ferromagnetism and superconductivity.}
\textbf{a}, $\Delta B$ as a function of $n_e$ and $D$ for a spatial position with sensitivity to both in- and out-of-plane magnetic moments (see Methods and Extended Data Fig. \ref{fig:SuppFigSpinCanting}), measured at $T=210$ mK and $B_z = 750 \mu$T.
\textbf{b}, Schematic phase diagram showing overlap of superconductivity (SC), the spin-canted ferromagnetism, and the valley-imbalanced ferromagnetism (VI). The dashed contour outlining the SC at 70 mK (light cyan region in Fig. \ref{fig:fig1}b) is shown as a guide to eye.
\textbf{c}, High-resolution linecut of $\Delta B$ along the trajectory marked in panel a, which is symmetrized and anti-symmetrized in $B_z = \pm 100 \mu$T. The antisymmetric contribution chiefly shows the Meissner response in the superconducting pocket, whereas the symmetric contribution highlights the in-plane moments (unaffected by the direction of $B_z$) coming from the ordered, spin-canted phase. 
\textbf{d}, Schematic of $m_\parallel$ and $m_\perp$ as a function of $n_e$, based off of the data in panel c. Insets show the spin -ordering of the $K$ (orange) and $K'$ (green) valley states, which are canted at higher $|n_e|$ and spin-valley locked at low $|n_e|$.}
\label{fig:fig_spincanting}
\end{figure}

\section{Superconductivity and spin-canted order}

Armed with a local magnetic probe of the superconducting state, we may correlate the onset of superconductivity with the magnetic phases that occur nearby in the phase diagram. 
Nearly all graphitic superconductors fall within or near a phase with broken isospin symmetry, and WSe$_2$-proximitized R3G is no exception\cite{patterson_superconductivity_2025,yang_impact_2025}.
The addition of WSe$_2$ to rhombohedral graphene layers enhances the strength of spin-orbit coupling on the proximal layer\cite{khoo_-demand_2017,island_spinorbit-driven_2019}.  Past experiments on bi- and trilayers found the emergence of additional superconducting pockets in the $n_e$-$D$ space\cite{zhang_enhanced_2023,zhang_twist-programmable_2025,li_tunable_2024,holleis_nematicity_2025,patterson_superconductivity_2025,yang_impact_2025}, typically with much higher $T_c$ that is found in the absence of the proximity enhanced spin-orbit coupling. While the mechanism for this enhancement is not known precisely \cite{chatterjee_inter-valley_2022,jimeno-pozo_superconductivity_2023}, one theoretical proposal attributes the enhanced superconductivity to the onset of `spin-canted' order, arising from the competition between ferromagnetic interlayer Hund's coupling and the enhanced spin orbit \cite{dong_superconductivity_2026,raines_superconductivity_2026}.  In this picture, superconductivity occurs in the vicinity of a continuous phase transition from a spin-valley locked to a spin canted phase, with pairing mediated by ferromagnetic fluctuations.  
Previous magnetometry experiments in the normal state of R3G/WSe$_2$ \cite{patterson_superconductivity_2025} indeed found evidence for this transition in the vicinity of the superconducting region of the  $n_e$-$D$ plane.  However, the lack of a local probe of superconductivity precluded precise determination of the coexistence and/or competition between superconducting and spin-canted order. 

To make a direct comparison between the onset of superconductivity and spin canted order, we leverage the difference in spatial structure of the magnetic fringe fields arising from in- and out-of-plane moments as well as their different responses to inversion of $B_z$ and $B_\parallel$
(see Methods and Extended Data Fig.\ref{fig:SuppFigSpinCanting}).
In Fig. \ref{fig:fig_spincanting}a, we present $\Delta B$ measured at $T= 210$ mK and $B_z = 750$ $\mu$T, measured at a point in the sample where fringe fields from in-plane moments are somewhat more pronounced.  
In addition to signals arising from the Meissner effect and the valley imbalanced (VI) orbital magnet described in Fig. \ref{fig:fig1}c, we 
observed a broad region of slightly negative $\Delta B$  adjacent to the superconducting region but at higher $D$. This region is coincident with where previous nSOT measurements\cite{patterson_superconductivity_2025} found a phase with in-plane magnetic anisotropy, associated with spin-canted order (see Fig. \ref{fig:fig_spincanting}b, where we show a schematic phase diagram).  

To characterize the relationship between superconductivity and spin canting precisely, we decouple spin-canting and Meissner signals by symmetrizing and anti-symmetrizing the measured $\Delta B$ with respect to changing the sign of the applied $B_z$, in a nearby location with greater sensitivity to in-plane moments. With fixed $B_x =-46$ mT, we define $\Delta B_{+(-)}=\frac{1}{2}(\Delta B(B_z =+0.1\rm{ mT}) \pm \Delta B(B_z =-0.1\rm{ mT})$, and plot these quantities along the dashed $n_e-D$ trajectory shown in Fig.\ref{fig:fig_spincanting}a in Fig. \ref{fig:fig_spincanting}c.
To within our experimental resolution, we find that the spin canting signal $\Delta B_+$ goes continuously to zero, consistent with a second order transition.  Superconductivity onsets at the edge of the spin canted phase, with the bulk of the superconducting region, and maximal $T_c$, observed on the high $n_e$ side of the transition where no spin-canted order is detected. 
This relationship is summarized in Fig.\ref{fig:fig_spincanting}d, where we show the relationship of the superconducting `dome' and the onset of spin canted order at $T=210$ mK. Notably, as shown in Fig. \ref{fig:fig_spincanting}b, at lower $T=70$ mK the superconducting state extends even deeper into the spin-canted state.  We thus conclude that superconductivity indeed straddles the quantum phase transition between the spin-canting and spin-valley locked states.

\begin{figure*}[ht!]
\centering
\includegraphics[width=15cm]{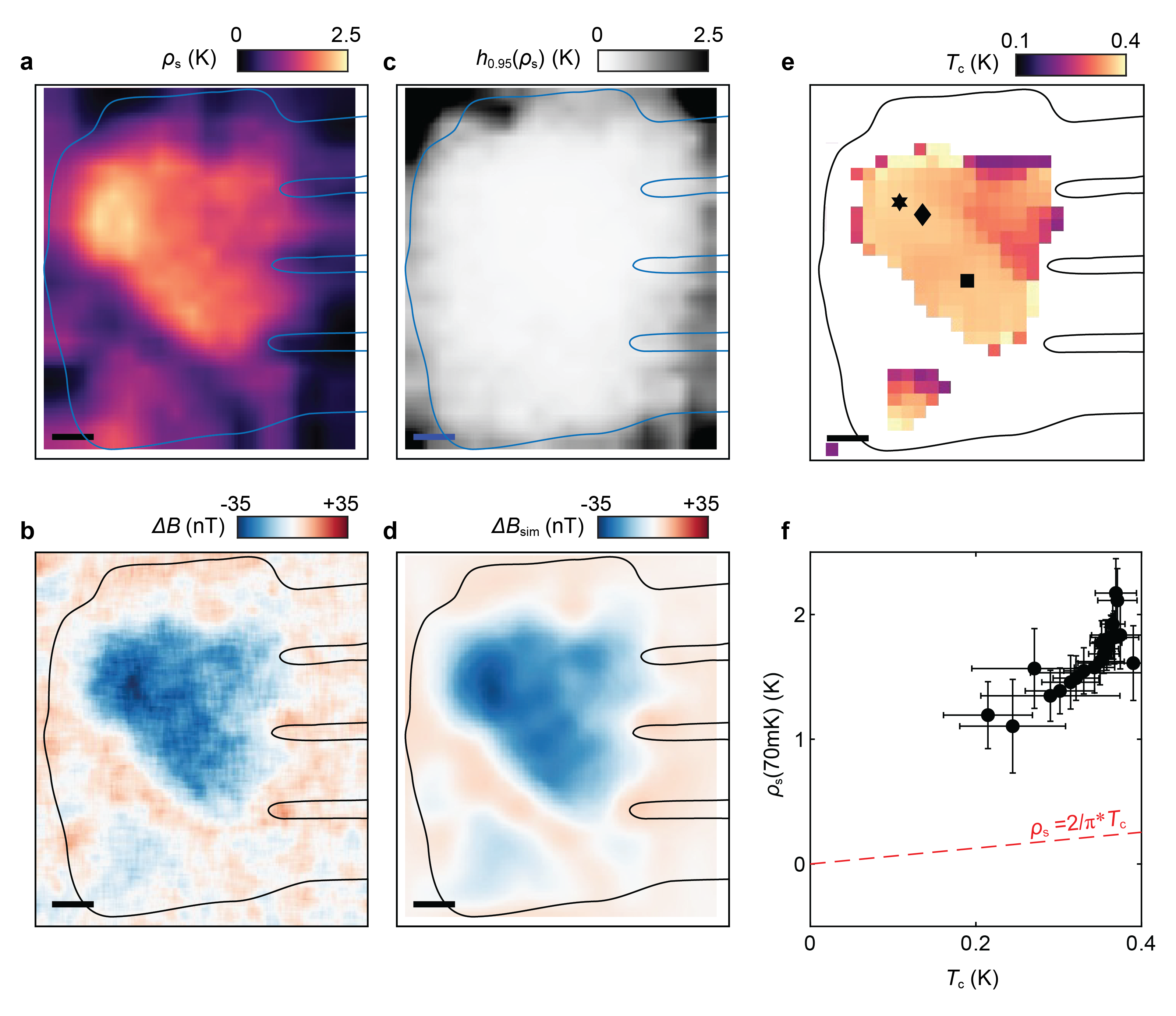}
\caption{\textbf{Quantifying the local superfluid stiffness $\rho_s$ and local $T_c$.}
\textbf{a}, Reconstructed spatial map $\rho_s$ from $\Delta B$ measured at $T = 70$ mK.
\textbf{b}, Experimental data of screening field $\Delta B$ in an external $B_z = 150 \rm{\mu T}$ used for the reconstruction in panel a.
\textbf{c}, Half-width of the 95\% confidence interval for the least-squares minimization presented in panel a.
\textbf{d}, The simulated screening field corresponding to the $\rho_s$ spatial map in panel a, which matches data in panel b.
\textbf{e}, Measurement of $T_c$ as a function of space, defined as the temperature at which the local $\Delta B$ goes to zero (Methods). Symbols represent the spatial locations of data in Fig. \ref{fig:fig_tdep}.
\textbf{f}, Uemura-style plot for correlating $\rho_s$ (panel a) and $T_c$ (panel e) pixel-by-pixel. The red dashed line corresponds to the lower bound set by BKT limit with $\rho_s = 2T_c/\pi$.
}
\label{fig:fig_rhos}
\end{figure*}

\section{Quantifying local superfluid stiffness}

The local fringe field $\Delta B_z$ can also be used to quantitatively determine local thermodynamic properties of the superconductor, particularly $\rho_s(T)$.  
In the absence of vortices, we are free to adopt a London gauge so that the phase-gradient term in Equation \ref{eq:GL_London} vanishes, and the current--phase relation reduces to $\mathbf{j}_s = -\frac{4\pi^2 k_B \rho_s}{\Phi_0^2}\,\mathbf{A}$.
To determine $\rho_s$, we iteratively solve the `forward problem' in which we map a trial $\rho_s(x,y)$ to the implied fringe field response $\Delta B_{sim}(x,y)$ and compare this to the measured $\Delta B$ to determine a best fit (see Methods). 
Fig. \ref{fig:fig_rhos}a shows the best fit $\rho_s$ resulting from this procedure for the $T=70$ mK experimental data shown in Fig. \ref{fig:fig_rhos}b (additional data are shown in Extended Data Fig. \ref{fig:all_n_dep}).
To determine the confidence interval for this fit, we take the pixel-by-pixel standard error from the covariance matrix of the least-squares fitting; the uncertainty in $\rho_s$ derived from half-width of the $95\%$ confidence intervals is shown in Fig. \ref{fig:fig_rhos}c. As a verification of the quality of this fitting, we show the simulated $\Delta B_{sim}(x,y)$ in Fig. \ref{fig:fig_rhos}d for easy comparison. 
% Somewhat surprisingly, we do not observe a depression in $\rho_s$ at the equilibrium positions of the vortices shown in Fig. \ref{fig:fig2} or much correlation between $\rho_s$ and the spatial structure of the valley imbalanced ferromagnet.  

Turning to the quantitative aspects of this analysis, we find that at $T=70$mK, $\rho_s$ is as high as $2.3$ K---approximately $6$ times larger than $T_c\approx 400$ mK determined from the global transport measurements of  Fig. \ref{fig:fig1}f.
Indeed, temperature-dependent measurements of $\Delta B$ allow us to determine the \textit{local} $T_c$ (see Extended Data Fig. \ref{fig:all_t_dep}-\ref{fig:tcmap_ana} and Methods), defined as the temperature where a finite Meissner signal is first observed (Fig. \ref{fig:fig_rhos}e). 
While the microscopic origin of the spatial structure of both $\rho_s$ and $T_c$ is uncertain, its existence allows us to examine the correlation between these two thermodynamic quantities. 
As shown in Fig. \ref{fig:fig_rhos}, $\rho_s$ and $T_c$ are correlated.  
The maximum local $T_c$ matches the measured transport $T_c$ of nearly $400mK$.  Moreover, $\rho_s$ is always considerably larger than local $T_c$, and thus the nominal phase fluctuation limit of $\rho_s = \frac{2}{\pi}T_c$.

\begin{figure*}[ht!]
\centering
\includegraphics[width=18cm]{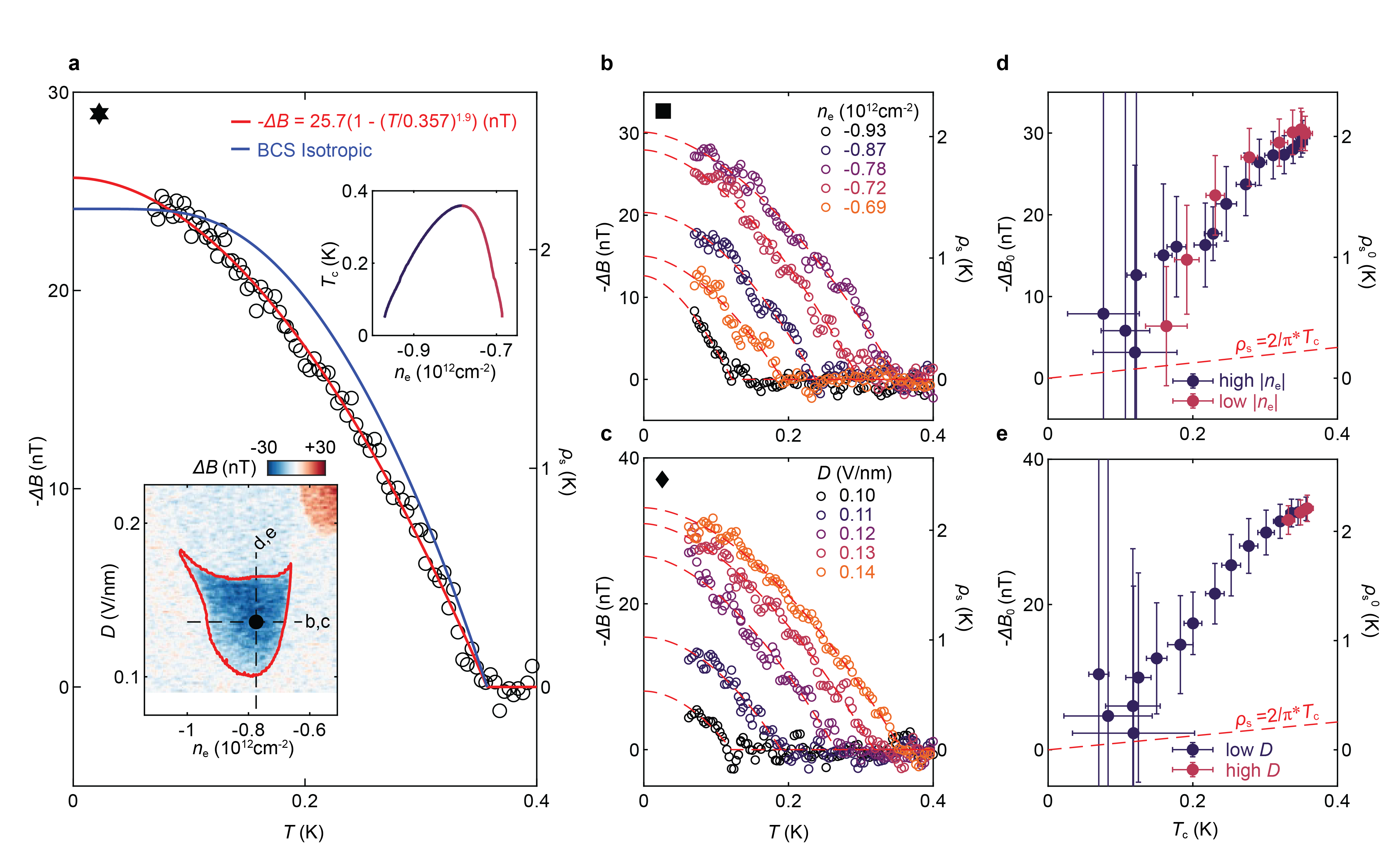}
\caption{\textbf{Temperature dependence of $\rho_s$ and Uemura plot.}
\textbf{a}, $\Delta B$ and $\rho_s$ measured at a single location as a function of $T$.
The data is taken at the $n_e$ and $D$ indicated by the solid dot in the bottom left inset, which shows $\Delta B$ phase diagram at $T=70$ mK. The upper right inset panel is the transport measurement determined $T_c$ along the dashed horizontal trajectory in the bottom left inset.
The red curve shows a power law fit, while the blue curve shows the expectation for an isotropic, $s$-wave BCS superconductor.
\textbf{b}, $\Delta B$ (or $\rho_s$) as a function of $T$ for various $n_e$ along the horizontal trajectory in the inset of panel a. Dashed lines show power law fits to the data.
\textbf{d}, Uemura plot of the $y$-intercept $\Delta B^0$ (or $\rho_s^0$) versus $T_c$, as determined in panel b. The color of the data points indicates whether $n_e$ is higher or lower relative to the value at optimal $T_c$.
\textbf{c, e}, Same as panel b, d but for various $D$ along the vertical dashed trajectory in the inset of panel a. 
}
\label{fig:fig_tdep}
\end{figure*}

\section{Temperature dependence of $\rho_s$}

The temperature dependence $\rho_s(T)$ below $T_c$ is determined by quasiparticle excitations and contains information about the gap structure.  
To study this in detail, we measure the local $\Delta B$ as a proxy for $\rho_s$.
As compared to acquiring the 2D images of $\Delta B$, from which $\rho_s$ is then determined by fitting, measuring $\Delta B$ at a single point allows for longer averaging and thus increases the signal to noise ratio.  
Measuring local $\Delta B$ also avoids systematic errors introduced by the fitting algorithm when $\Delta B$ is close to our noise floor.  
Specifically, when the signal to noise ratio is poor, the algorithm will attribute noise signals to unphysical diamagnetic screening currents, leading to an overestimate of $\rho_s$. 
Conversely, the nature of fringe magnetic fields is such that local $\Delta B$ is not always proportional to the local $\rho_s$.  As a concrete example, fringe fields just outside the boundary of a superconducting region will be opposite in sign to those in the bulk.  
To mitigate this concern, we first measure $\rho_s(T)$ (via 2D imaging) and then focus on locations where $\rho_s(x,y)\propto \Delta B(x,y)$ (see Methods); most obviously, this condition is met where $\rho(x,y)$ is at a (spatial) maximum. 
Focusing on these locations for ultra-high averaging local measurements of $\Delta B(T)$, we are able to measure a qualitatively accurate trend.  We then fix the absolute scale factor between local $\rho_s$ and local $\Delta B$ from a two dimensional image taken at the lowest temperatures where the signal-to-noise is highest and systematic errors introduced by fitting algorithm can be neglected. 

Fig. \ref{fig:fig_tdep}a shows $\rho_s(T)$ from one such measurement.  Diamagnetism onsets at a local $T_c \approx 350$ mK. Qualitatively similar data is observed in several sample locations (Extended Data Fig. \ref{fig:tcmap_ana}).  
In a pairing limited superconductor, $\rho_s(T)$ (or more precisely, $\rho_s(T=0)-\rho_s(T)$) measures the density of thermally excited quasiparticles, constraining the structure of the superconducting gap $\Delta$ \cite{prozorov_magnetic_2006}.
For example, in a clean s-wave superconductor, the quasiparticle density follows an Arrhenius law with $(\rho_s(T=0) - \rho_s(T))\propto T^{-1/2}e^{-\Delta(0)/(k_B T)}$, where $\Delta(0)=1.76 k_B T_c$ is the superconducting gap at zero T=0. At another extreme, in a clean nodal superconductor, quasiparticles are excited even at deeply sub-gap temperatures, leading to a linear $\rho_s(T)\propto 1-T$. 
Our data are not well fit by an isotropic $s$-wave superconductor \cite{tanaka_superfluid_2025}, nor do they show a clear linear-in-$T$ dependence.  Rather, data are well fit by a power law, $\rho_s (T) = \rho_s^0(1-(T/T_c)^n)$, where the zero temperature stiffness $\rho_s^0$, $T_c$ and $n$ are free parameters and the typical best fit power $n\approx 1.9$. 
This temperature dependence could be consistent with an anisotropic gap \cite{tanaka_superfluid_2025} or distinct gaps on multiple bands  \cite{manzano_exponential_2002,kogan_superfluid_2009,luan_local_2010}.

Applying the same power law fit to data taken at different $n_e$ (Fig. \ref{fig:fig_tdep}b) and $D$ (Fig. \ref{fig:fig_tdep}c) allows us to determine the relation between $T_c$ and $\rho_s^0$ across the superconducting pocket shown in the inset of Fig. \ref{fig:fig_tdep}a.  
These results are summarized in the Uemura plots \cite{uemura_universal_1989} of Figs. \ref{fig:fig_tdep}d and e.  $\rho_s^0$ and $T_c$ are found to be linearly correlated, with $\rho_s^0\approx 5.5T_c$ for data taken at the same position but varying $n_e$ and $D$, consistent with the correlation found as a function of position for fixed $n_e$ and $D$ in Fig. \ref{fig:fig_rhos}f.  

\section{Discussion} 

The technical advances reported here open the path to thermodynamic characterizations of ultra-low density superconductors across a wide range of 2D materials and heterostructures.  As shown for this particular sample, this approach can help resolve key questions about real- and parameter-space correlations between different physical phenomena.  
One immediate surprise revealed by the current experiment is the degree of mesoscopic disorder present in the sample on few-micron length scales.  Notably, there is a clear discrepancy in real-space distribution of ferromagnetic moments in the valley imbalanced phases (Fig. \ref{fig:fig1}d) and Meissner screening in the superconducting phase (Fig. \ref{fig:fig1}e)---while many parts of the sample show both superconductivity and ferromagnetism (at the $n_e,D$ coordinates appropriate for each phase), in other regions one of the two phenomena is significantly suppressed. This behavior may be tied, in the current sample, to the presence of both a WSe$_2$ and an aligned hBN cladding layer.  Prior experiments on WSe$_2$ clad R3G devices with a misaligned hBN\cite{patterson_superconductivity_2025,yang_impact_2025} show superconductivity very similar in extent and properties to the sample studied here; conversely, hBN alignment is known to significantly alter the extent of some of the ferromagnetic phases\cite{uzan_hbn_2025}.  Misalignments or contamination of the two interfaces in the current device may thus have disparate effects on superconductivity and ferromagnetism.  This question may be resolved in future studies of hBN-clad superconducting rhombohedral multilayers in which the substrate and graphene crystal axes are misaligned, so that the correlated electron physics is not sensitive to substrate coupling.

As noted above, our data is qualitatively consistent with a proposed mechanism for superconductivity mediated by spin-canting fluctuations \cite{dong_superconductivity_2026,raines_superconductivity_2026}. In particular, we identify a critical point at which the spin-canting order vanishes continuously within the superconducting dome (Figs. \ref{fig:fig_spincanting}c-d). Theoretically, the pairing strength is expected to be stronger in the ordered phase, but we observe that $T_c$ is maximal on the disordered side of the phase transition.
This may be explained by the near-coincidence of the spin-canting transition with a Lifshitz transition.  Throughout most of the spin canted phase, the electron system is a simple half metal, with two Fermi pockets. However, in most of the spin valley locked phase, the Fermi surface topology includes minority pockets \cite{patterson_superconductivity_2025,yang_impact_2025}. These asymmetric pockets are thought to play a key role in some theories for magnon-mediated pairing \cite{dong_superconductivity_2026} while other theories find similar superconducting instabilities without minority pockets \cite{raines_superconductivity_2026}. Our observation of a superconducting dome centered over the multi-band spin-valley locked phase suggests that the details of inter-band scattering may be the dominant source of tunability of $T_c$ with gate voltage. 

The description of the superconducting state itself revealed by our measurements can be summarized by two key facts. First, the temperature dependence of $\rho_s$ resembles that of a weak-to-intermediate coupling superconductor: $\rho_s(T)\propto T-T_C$ close to $T_c$ as expected from mean field theory, while $\rho_s^0\approx 5.5 T_c$, placing it well outside the strong phase fluctuation regime\cite{emery_importance_1995}.  Second, $\rho_s^0\propto T_c$ over a wide range of $T_c$ as tuned by experimental parameters such at $n_e$ and $D$.   Such proportionality is not expected in clean BCS superconductors, in which $\rho_s^0$ is set by the effective Fermi energy $E_F$. 

$\rho_s^0\propto T_c$ does arise in the `dirty-limit' of BCS superconductivity, when the coherence length $\xi$ is much larger than the electron mean free path $\ell_{MF}$ \cite{tinkham_introduction_2015}.
However, this limit is very unlikely to apply to rhombohedral graphene \cite{zhou_superconductivity_2021, seo_family_2025}.  
In the current sample, the upper critical field is found to be $B_z\approx 50 mT$, which sets an upper bound on of $\xi\lesssim 80 \rm{nm}$.  The mean free path can be estimated from the normal state resistivity $R_{\square}\approx 200 \Omega$ as $\ell_{MF}=\frac{\Phi_0}{4R_{\square}k_F}\approx 500\rm{nm}$.  Alternatively, $\ell_{MF}$ can be bounded from the magnetic field where quantum oscillations onset, $B_z\approx 500 mT$ \cite{patterson_superconductivity_2025,yang_impact_2025}, at which point $\ell_{MF}$ should be much larger than the classical cyclotron radius $R_C\approx k_F\ell_B^2$ (where $\ell_B$ is the magnetic length).  
This estimate gives a consistent value of $\ell_{MF}\gtrsim 200nm$, again larger than $\xi$.  Our sample is thus deep in the clean limit, defined as $\xi\lesssim\ell_{MF}$. 

$\rho_s^0 \propto T_c$ is also expected when phase fluctuations determine $T_c$\cite{emery_importance_1995}.  However, this picture implies a near-unity constant of proportionality, $\rho_s^0/T_c\sim 1$.  This ratio is observed in cuprate superconductors\cite{uemura_universal_1989,bozovic_dependence_2016} but is violated in the present experiment on R3G/WSe$_2$. 
Explaining the fact that $\rho_s^0\propto T_c$ with a proportionality constant considerably larger than one in the absence of disorder is an outstanding theoretical problem. 
Further investigation may reveal whether this observation is generic to the wide variety of superconducting states observed in different graphene allotropes\cite{cao_unconventional_2018,zhou_superconductivity_2021,han_signatures_2025} and related two-dimensional materials\cite{guo_superconductivity_2025,xia_superconductivity_2025,xu_signatures_2025}.

%\bibliographystyle{custom}
%\bibliography{references,references_2}
%%BIB START

%BIB END
% Methods
\section*{Methods}

\subsection*{Sample fabrication}
All data is presented from a single device, which was also described (as Device E) in Ref. \cite{patterson_superconductivity_2025}. In brief, the rhombohedral trilayer graphene (R3G) flake is identified using Raman spectroscopy and isolated via local anodic oxidation with an atomic force microscope. Separately, a poly(bisphenol A carbonate) stamp is used to pick up an hBN/few layer graphite stack, which is then dissolved in chloroform and vacuum annealed to remove the polymer. Then, a second stamp is used to pick up a few-layer-graphite/hBN/monolayer WSe$_2$/R3G stack. This stack is dropped onto the first half, such that the R3G is aligned to the bottom hBN, forming a moire pattern. The device is then etched and contacted using standard electron beam lithography techniques. An optical micrograph and atomic force microscopy image of the completed device are presented in Extended Data Fig. \ref{fig:PhaseDiagrams}. 

\subsection*{NanoSQUID on tip measurements}

The nanoSQUID on tip (nSOT) sensor was fabricated by cryogenic evaporation of indium onto the apex of a pulled quartz pipette. The nSOT is measured via a feedback loop using a series SQUID array amplifier, and has a sensitivity of $\sim 1-3\rm{nT}/\rm{Hz}^{1/2}$ at measurement frequencies (circuit shown in Extended Data Fig. \ref{fig:nSOTChar}a). In order to calibrate the response of the SQUID from current through the feedback circuit into flux through the nSOT, we step the external magnetic field by 100 $\mu$T before and after measurements to establish a linear conversion factor between measured current and applied flux. From both SEM imaging and the field-period of the SQUID, we estimate the size of the nSOT to be 350 nm diameter, which provides the conversion between $\Delta B$, magnetic field through the nSOT, and $\Delta \Phi$, magnetic flux through the nSOT, presented in the main text. The nSOT is positioned over the sample at a distace of $150 - 200$ nm with a Attocube ANSxyz100std/LT xyz-Scanner. A quartz tuning fork attached to the nSOT is operated in a phase-locked loop, which enables periodic measurement of the sample surface to ensure an approximately constant height. nSOT measurements are performed in a custom, wet dilution refrigerator in two modalities. In one modality, a small amount of Helium-4 exchange gas is left in the cell containing the microscope, which provides better thermalization/stabilization of the nSOT and tuning fork, at the cost of additional heating of the sample from the nSOT (see discussion below). Some measurements are performed with the sample and microscope in vacuum, after pumping out the sample cell, which results in lower sample temperatures.

In order to maintain high sensitivity of the nSOT at zero perpendicular magnetic field, which is crucial for studying the superconducting state, we mount the nSOT sensor at $\theta \approx 12.5^\circ$ off from the vertical axis of the microscope and apply an in-plane magnetic field $B_x = -46$ mT to tune the nSOT to a sensitive working point (Extended Data Fig. \ref{fig:nSOTChar}b). With this constant $B_x$, the nSOT is sensitive across a few mT of applied $B_z$. Due to a finite, inadvertent tilt $\alpha$ of the sample in the plane (Extended Data Fig. \ref{fig:nSOTChar}b), the in-plane magnetic field induces a small perpendicular magnetic field on the sample, which needs to be canceled out with the external field $B_z$. We find that the perpendicular magnetic field on the sample is zero when the $B_z = 1.9$ mT on the external SC magnet, indicating an approximately $\alpha= 2^\circ$ tilt of the sample. This is measured via the field at which the Meissner screening field is zero, as well as independent transport measurements which show the extent of the SC state maximized at this field. For all measurements presented in the main text, we subtract off this offset for clarity, so that $B_z$ expressed in the main text represents the approximate $B_{\perp}$ of the sample, rather than the applied magnetic field of the external magnet. 

Measurements of $\Delta B$ are performed using a quasi-DC boxcar technique in which the gates are oscillated using a square-wave signal between the gate voltage of interest and a non-magnetic reference point. In order to minimize parasitic electric field contrast from the top-gate, only the back-gate is oscillated at finite frequency, and the magnitude of the ``step" is minimized as much as possible while ensuring nonmagnetic reference points. For all data in the main text, we subtract out an additional ``background" signal to remove this parasitic electric field contribution by subtracting a separate measurement at the same location with the same voltage ``step" between two non-magnetic points in parameter space. 

The square wave applied to the back-gate and the expected signal from the nSOT is illustrated in Extended Data Fig. \ref{fig:Boxcar}. To suppress displacement currents caused by abrupt gate voltage steps, we inserted analog RC filters (cutoff frequency $f=5~\text{kHz}$, 48dB/octave roll off) on both the top- and bottom-gate lines (Extended Data Fig. \ref{fig:Boxcar}a). The combination of line resistance, parasitic capacitance, and the multi-pole cold RC filters on the transport contacts introduces a finite settling time, leading to a delayed saturation in the nSOT signal (Extended Data Fig. \ref{fig:Boxcar}b). To accurately extract the steady-state response, we used a boxcar averager that demodulates the nSOT signal by integrating over a gate window aligned with the saturation portion of the response. A baseline window of equal duration is subtracted from the gated signal to yield the static nSOT response.

The boxcar averaging was implemented using a Zurich Instrument MFLI digital lock-in amplifier. Because the saturated portion of the nSOT signal occupies only a small duty cycle, the effective averaging time is limited, often resulting in a lower signal-to-noise ratio in the boxcar output. To improve noise performance, we simultaneously demodulated the same input signal using the standard lock-in technique on the same instrument. The lock-in approach averages over the entire duty cycle, thereby suppressing noise more effectively, and yields a signal proportional to the static nSOT response by a constant factor. In all reported data, we calibrated this proportionality factor between the boxcar and lock-in results, and present the corrected lock-in signal as the final data.

\subsection*{Sensing in-plane and out-of-plane moments with nSOT}

The measured $\Delta B$ can include contributions from both in-plane and out-of-plane magnetization. While a detailed discussion can be found in \cite{patterson_superconductivity_2025}, we present further data supporting this in Extended Data Fig. \ref{fig:SuppFigSpinCanting}. Extended Data Fig. \ref{fig:SuppFigSpinCanting}a-b show schematics of the expected magnetic fringe field around a nonactive region of the sample, which could be a physical `hole' or disorder-induced reduction of magnetism. The nSOT measurement of $\Delta B$, which is chiefly in the $z$-direction, has a sign change in the center of the hole for out-of-plane oriented moments vs. a dipole-like behavior for in-plane oriented moments. In Extended Data Fig. \ref{fig:SuppFigSpinCanting}c-d, we present images with the device tuned into an orbital ferromagnetic state with out-of-plane magnetization and a spin-canted state with in-plane magnetization, respectively. Both images are taken with $B_z = 250 \mu T$ out of the plane of the sample, and $B_x = 46$ mT at a direction approximately 20$^\circ$ off of the vertical axis of our spatial images--i.e., the dipole seen in Fig. \ref{fig:SuppFigSpinCanting}d is oriented along the $B_x$ direction. Upon flipping the sign of $B_x$, the direction of the spin-canting dipole should flip while the orbital ferromagnet would be unchanged. Alternatively, reversing $B_z$ should reverse the direction of the orbital ferromagnet but not affect the spin-canted phase (in-plane moment). In Extended Data Fig. \ref{fig:SuppFigSpinCanting}e-g, we show a more complete dataset of the latter approach, as $B_z$ is tuned continuously across a trajectory in $n_e$-$D$ plane that goes from the spin-canted phase (which is broadly seen to be $B_z$ insensitive) into the superconductor (which has a sign reversal as $B_z$ crosses zero). The data presented in Fig. \ref{fig:fig_spincanting} in the main text comes from symmetrizing and anti-symmetrizing this dataset at $B_z = \pm 0.1$ mT; for completeness we show similar analyses at $B_z = \pm 0.05,0.1,0.15,0.2$ mT in Extended Data Fig. \ref{fig:SuppFigSpinCanting}g.

\subsection*{Extracting superfluid stiffness from measured screening field}
We extracted local 2D superfluid stiffness $\rho_{s}\left(x, y\right)$ by quantitatively fitting the experimental screening field result to forward simulations performed with open-source finite-element package SuperScreen \cite{bishop-van-horn_superscreen_2022}, which solves the 2D London equation for thin-film superconductors of arbitrary geometry and spatially varying stiffness. We first defined a coarse grid that covered the active region together with a surrounding padding area of nominally zero screening field, and represented the $\rho_s$ on this grid. We then bilinearly interpolated to the full resolution of the spatial image prior to each forward calculation. The forward model takes the applied $B_z$ and trial $\rho_s$ map as inputs and returns the vector magnetic field at a given height. Considering the tip-sample distance 250nm and tip tilt, we computed the expected $\Delta B(x,y)$ image. We then minimized a nonlinear least-squares objective with respect to the coarse-grid $\rho_s$, using the difference between simulated and measured screening fields as the primary residual $r_0 = B_{meas} - B_{sim}$, and imposed Tikhonov-type regularization terms on both the spatial gradients of $\rho_s$ and the gradients of the magnetic field, $r_1 \propto \nabla \rho_s$, $r_2 \propto \nabla^2 \rho_s$, and $r_3 \propto \nabla B_{meas} - \nabla B_{sim}$, to favor smooth yet sharply bounded stiffness profiles. Three scalar regularization parameters $\lambda_1$, $\lambda_2$, and $\lambda_3$ controlled the relative weights of these terms and were tuned on each dataset to obtain $\rho_s$ solutions that were smooth inside the superconducting region while preserving the amplitude.

Formally, we treated the discretized stiffness values on the coarse grid as the parameter vector $\mathbf{\rho}$. The fit minimizes the nonlinear least-squares cost
$$S(\mathbf{\rho})=\sum_{i=1}^{N} r_i(\rho)^2,$$
where $r(\mathbf{\rho})$ collects the residual $r_0$ and the regularization terms $r_1, r_2, r_3$. We considered only the residual $r_0$ for the confidence interval. Near the optimum $\hat{\rho}$, we linearize the residuals as
$$r(\rho)\approx r(\hat{\rho}) + J(\hat{\rho})(\rho-\hat{\rho}),$$
with Jacobian
$$J_{ij} = \frac{\partial r_i}{\partial \rho_j}|_{\rho=\hat{\rho}},$$
so each column of $J$ describes how the residual changes when perturbing the $\rho_s$ at a given pixel. Under the usual assumption of independence, homoscedastic Gaussian noise, this linearization yields and approximate covariance matrix for the fitted $\rho_s$ matrix,
$$Cov(\hat{\rho})\approx s^2(J^TJ)^{-1},\, s^2=\frac{1}{N-p}\sum_{i=1}^{N}r_i(\hat{\rho})^2,$$ where $N$ is the number of data points and $p$ is the dimension of the coarse $\rho_s$ matrix. The variance of the fitted stiffness at pixel $j$ is $[Cov(\hat{\rho})]_{jj}$ giving a standard error in physical unit Kelvin.
$$SE(\hat{\rho_j})=\sqrt{[Cov(\hat{\rho})]_{jj}}.$$

In addition to Fig.\ref{fig:fig_rhos}, we present another representative $\rho_s$ fit in Extended Data Fig.\ref{fig:all_n_dep}a. Notably, although no explicit boundary conditions were imposed on the supercurrent distribution, the reconstructed current magnitude $||j||$ remains naturally confined to the active region of the device. We attribute this self-confinement to the high spatial resolution of the nSOT.

\subsection*{Extracting local $T_c$ from measured screening field}

To extract local $T_c$, we acquired a temperature series of screening-field images under a fixed out-of-plane field $B_z = 150\mu T$ (Extended Data Fig.\ref{fig:all_t_dep}). At each pixel, we compiled the temperature dependence into a single $\Delta B(T)$ trace and fit this trace to a power-law form $\Delta B(T)=A[1-(T/T_c)^n]$ with a piecewise termination: $\Delta B=0$ when $T>T_c$, using standard nonlinear least-squares regression with bounded parameters. As an initial estimate of $T_{c,\text{ext}}$, we identified the lowest temperature at which the measured screening field exceed $-3nT$; if no such crossing was found at a given pixel, we assigned a default initial estimate $T_{c,\text{ext}}=20mK$ for that location. The fit was constrained to $A\leq0$, $0\leq n\leq5$, and $T_c \geq 0$. This procedure yields, for every spatial pixel, a best-fit critical temperature $T_c$ together with confidence intervals obtained from the covariance of the least-squares fit. For pixels with no valid fitting, the confidence intervals are invalid, thus the corresponding $T_{c,\text{local}}$ is left blank.

\subsection{Linear approximation between $\rho_s$ and $\Delta B$}

As discussed and presented in the main text, at certain well-behaved locations, we adopted an approximate linear relation between the measurement of $\Delta B$ and the extracted $\rho_s$ from the magnetic inversion described above.

Here, we describe an analytical model for why this approximation is reasonable near smooth regions of the superfluid stiffness (i.e., where $\nabla \rho_s$ is small) and in the weak-screening limit.

In the 2D limit, the inhomogeneous London equation is
\begin{align}
\label{eq:inhomo_london_2d}
\mathbf{B}
&= -\nabla \times \left(c \, \frac{\mathbf{j}_s}{\rho_s} \right) \\
&= -c\left( \frac{1}{\rho_s}\,\nabla \times \mathbf{j}_s - \frac{1}{\rho_s^2} \nabla \rho_s \times\mathbf{j}_s  \right)
\end{align}
with $c=\frac{\Phi_0^2}{2\pi^2k_B}$ being the prefactor \cite{bishop-van-horn_superscreen_2022,cave_critical_1986,kogan_meissner_2011}.
By choosing a spatial location with $\rho_s$ being roughly constant nearby, the second term vanishes. 
Meanwhile, the weak-screening limit guarantees that the left-hand side of this equation is approximately the externally applied magnetic field $B_z$, so we have 
\begin{align}
    \rho_s\approx  -\frac{c}{B_z} \nabla\times \mathbf{j}_s
    \label{eq:approxrhos}
\end{align}

The measured $\Delta B$ by the nSOT can be thought of as generated by the Meissner screening current $j_s$ in the sample: in a thin film, the out--of--plane magnetic field at sensor height $h$ is related to the in--plane sheet current density $\mathbf{j}_s$ via the Biot--Savart law. In Fourier space, this relation can be written as
\begin{equation}
\Delta B_z(\mathbf{k}, h) = \mu_0\, K(\mathbf{k}, h)\,\big[ \mathbf{k} \times \mathbf{j}_s(\mathbf{k})\big],
\end{equation}
where $\mathbf{k}$ is the in--plane wave-vector, $\mathbf{j}_s(\mathbf{k})$ is the Fourier transform of the in--plane screening current density, and $K(\mathbf{k}, h)$ is the scalar transfer function encoding the Biot--Savart kernel\cite{meltzer_direct_2017, zuber_new_2018}. For a film in free space one has $K(\mathbf{k}, h) \propto e^{-|\mathbf{k}| h}$, so each spatial Fourier component of $\mathbf{j}_s$ is multiplied by an exponential factor $e^{-|\mathbf{k}| h}$: current variations with $|\mathbf{k}| h \ll 1$ are transmitted essentially unchanged to $B_z$, whereas modes with $|\mathbf{k}| \gtrsim 1/h$ are strongly suppressed. If the screening currents vary only on length scales $L$ much larger than the sensor scales, $L \gg \max(h, R_{\mathrm{tip}})$, then all significant Fourier components satisfy $|\mathbf{k}| h \ll 1$, implying $e^{-|\mathbf{k}| h} \approx 1$ and hence
\begin{equation}
\Delta B_z(\mathbf{r}, h) \approx \text{const} \, \nabla \times \mathbf{j}_s(\mathbf{r}),
\label{eq:approxcurrent}
\end{equation}
on the scale we probe. Projected along the axis of the tip, we can combine these two Eqs. \ref{eq:approxrhos} and \ref{eq:approxcurrent} and find that
\begin{equation}
\rho_s\propto -\frac{\Delta B}{B_z}. 
\end{equation}

We present additional data in Extended Data Fig. \ref{fig:all_n_dep} and Extended Data Fig. \ref{fig:lin_t_dep} to support this approximation.
In Extended Data Fig. \ref{fig:all_n_dep}b, we vary the carrier density $n_e$ at $T=210mK$, which lies near the midpoint of the temperature range explored in our experiment. At three locations, we took a high-resolution $\Delta B$ sweep and overlay the $\rho_s$ fit result from seven spatial $\Delta B$ images. We observed a qualitative correlation in which $\rho_s$ and $-\Delta B$ follow similar trends. When the $-\Delta B$ signal is large relative to the experimental noise floor, the confidence interval of $\rho_s$ is correspondingly narrow. In contrast, when the $-\Delta B$ signal is weak, the uncertainty in $\rho_s$ increases significantly, in some cases exceeding the fitted value of $\rho_s$.
In Extended Data Fig. \ref{fig:lin_t_dep}, we extract the local $\Delta B$ at four locations and correlate the ratio of $-\rho_s/\Delta B$ with $\Delta B$ obtained from spatial images over the full temperature range (Extended Data Fig. \ref{fig:all_t_dep}). 
We find that the ratio $-\rho_s/\Delta B$ remains approximately constant when the $\Delta B$ signal is large and the fit uncertainty is small, and systematically deviates from this constant value as the fitting error becomes significant.

\subsection*{Characterization of tip-induced heating}

Because the nSOT is operated in the voltage state, there is constant heat dissipation at the tip, which can lead to heating of the sample. This effect is much more dramatic when there is exchange gas present in the microscope cell, as discussed above. In order to measure this effect, we use the superconducting transport of the device as a secondary thermometer. In Extended Data Fig. \ref{fig:TipHeating}a-b, we show $R_{xx}$ vs. $n$ at varying mixing chamber temperature $T_{MC}$ for a fixed displacement field in the sample. This map is taken with the nSOT grounded, such that there is no heating from the tip. We then measured the same $R_{xx}$ trace with the tip biased to the measurement configuration and at varying heights, down to the measurement height (Extended Data Fig.\ref{fig:TipHeating}c-d). Importantly, we do not observe widening of the superconducting transition (as a function of $n$) when the tip is biased, and within our experimental resolution, transport curves measured with nSOT-induced heating match well to measurements at elevated mixing chamber temperatures with no tip heating. This indicates that the tip uniformly heats the sample over at least the lengthscales over which the transport contacts probe. Temperatures reported in the main text are calibrated by matching the width (in $n_e$) of the superconductor with the tip biased to the same measurement with no tip heating, and thus describe the electron temperature of the sample accounting for heating from the tip. 

\section*{Acknowledgments} 
We acknowledge discussions with Étienne Lantagne-Hurtubise, Zhiyu Dong, and Steve Kivelson. 
AFY acknowledges primary support for this work by the Department of Energy under award DE-SC0020043 and the Gordon and Betty Moore Foundation's Experimental Physics Investigator award GBMF13801 who provided equal support. 
AFY acknowledges further support by the  W. M. Keck Foundation under award SB190132 for the development of the ultra-low temperature SQUID microscopy setup. 
KW. and TT acknowledge support from the JSPS KAKENHI (Grant Numbers 21H05233 and 23H02052), the CREST (JPMJCR24A5), JST and World Premier International Research Center Initiative (WPI), MEXT, Japan. 
TA, OS, and  ER acknowledge direct support by the National Science Foundation through Enabling Quantum Leap: Convergent Accelerated Discovery Foundries for Quantum Materials Science, Engineering and Information (Q-AMASE-i) award number DMR-1906325; the work also made use of shared equipment sponsored under this award. 
CJ acknowledges support from the Department of Energy under award DE-SC0026348.
Use was made of computational facilities purchased with funds from the National Science Foundation (CNS-1725797) and administered by the Center for Scientific Computing (CSC). 
The CSC is supported by the California NanoSystems Institute and the Materials Research Science and Engineering Center (MRSEC; NSF DMR 2308708) at UC Santa Barbara.
E.B. was supported by the Simons Foundation Collaboration on New Frontiers in Superconductivity (Grant SFI-MPS-NFS-00006741-03), by an NSF-BSF Award No. DMR-2310312, and by CRC 183 of the Deutsche Forschungsgemeinschaft (Project C02).
Some of this work was initiated at the Kavli Institute for Theoretical Physics, which is supported by grant NSF PHY-2309135.

\section*{Author contributions}
RZ and BAF performed the nSOT measurements, with assistance from OS and TA. EB provided theoretical support. RZ built the ultra-low temperature nSOT microscope, with assistance from OS, TA, and SG. SX and TX fabricated the device, under supervision of CJ. YG and HS performed additional transport characterization. BAF fabricated and characterized the nSOT, with assistance from AK, ER, and CZ. TT and KW provided hBN crystals. MEH provided SQUID array amplifiers for nSOT readout. RZ, BAF, OS, and AFY analyzed the data and wrote the paper with inputs from all authors.

\section*{Competing interests}
The authors declare no competing interests.

\section*{Data availability}
Source data that reproduce the plots are provided with this paper. All supporting data for this paper and other findings of this study are available from the corresponding authors upon request. 

\section*{Code availability}
All numerical codes for this paper and other findings of this study are available from the corresponding authors upon request.

% Extended Data Figures

\setcounter{figure}{0}
\renewcommand{\figurename}{\textbf{Extended Data Fig.}}

\begin{figure*}[ht!]
\centering
\includegraphics[width=18cm]{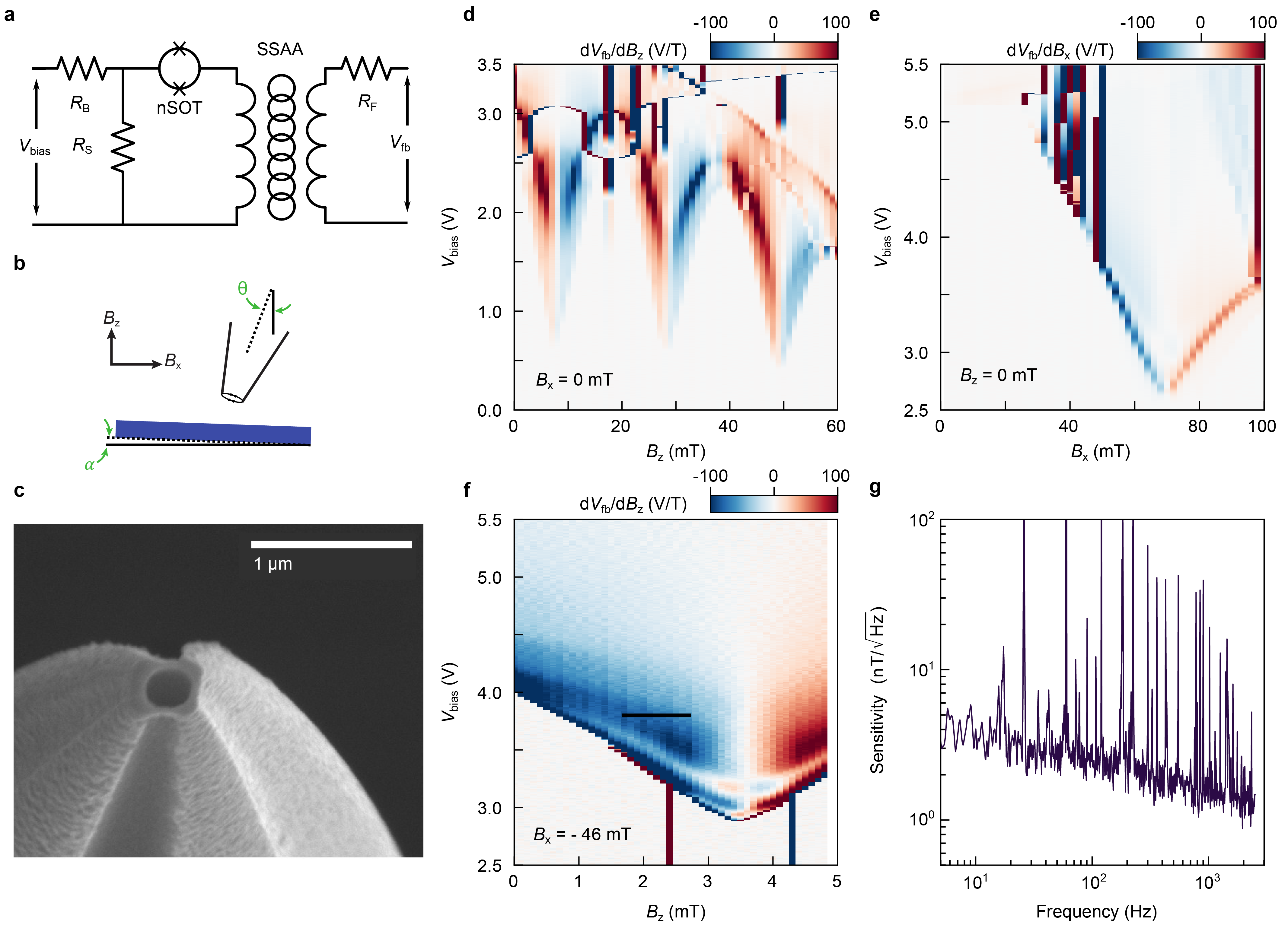}
\caption{\textbf{nanoSQUID-on-tip (nSOT) characterization.}
\textbf{a}, Schematic of nSOT measurement circuit. A voltage bias $V_{\rm{bias}}$ is applied across a bias resistor $R_B=10$ k$\Omega$, and a shunt resistor of $R_s = 5 \Omega$ shunts the nSOT. The current through the nSOT is measured via the voltage $V_{fb}$ across a resistor $R_F = 1$ k$\Omega$ coupled via a series superconducting squid array amplifier with a turn-ratio of 13.5 between the inductors.
\textbf{b}, Schematic of the nSOT in the $x-z$ plane of the external magnet. The nSOT is tilted at angle $\theta\approx 12.5^\circ$ relative to the vertical. There is also a small tilt $\alpha \approx 2^\circ$ due to imperfect mounting of the sample, causing an in-plane field $B_x$ to inadvertently apply a finite effective perpendicular field to the sample.
\textbf{c}, SEM image of a representative nSOT similar to the one used in this experiment.
\textbf{d}, Measured d$V_{fb}$/d$B_z$ of the nSOT as a function of applied bias voltage $V_{\rm{bias}}$ to the nSOT measurement circuit and applied magnetic field in the $z$-direction, $B_z$, at zero in-plane magnetic field, $B_x$. Jumps at high bias are artifacts due to the feedback circuit.
\textbf{e}, Measured d$V_{fb}$/d$B_x$ of the nSOT as a function of $V_{\rm{bias}}$ and $B_x$ at $B_z = 0$; the longer period of oscillation is due to the tilt of the nSOT. 
\textbf{f}, Measured d$V_{fb}$/d$B_z$ of the nSOT as a function of $V_{\rm{bias}}$ and $B_z$ at fixed $B_x = -46$ mT. The black bar denotes the bias and range of magnetic fields for the measurements. 
\textbf{g}, Representative sensitivity of the nSOT as a function of frequency.
}
\label{fig:nSOTChar} 
\end{figure*}

\begin{figure*}
\centering
\includegraphics[width=9cm]{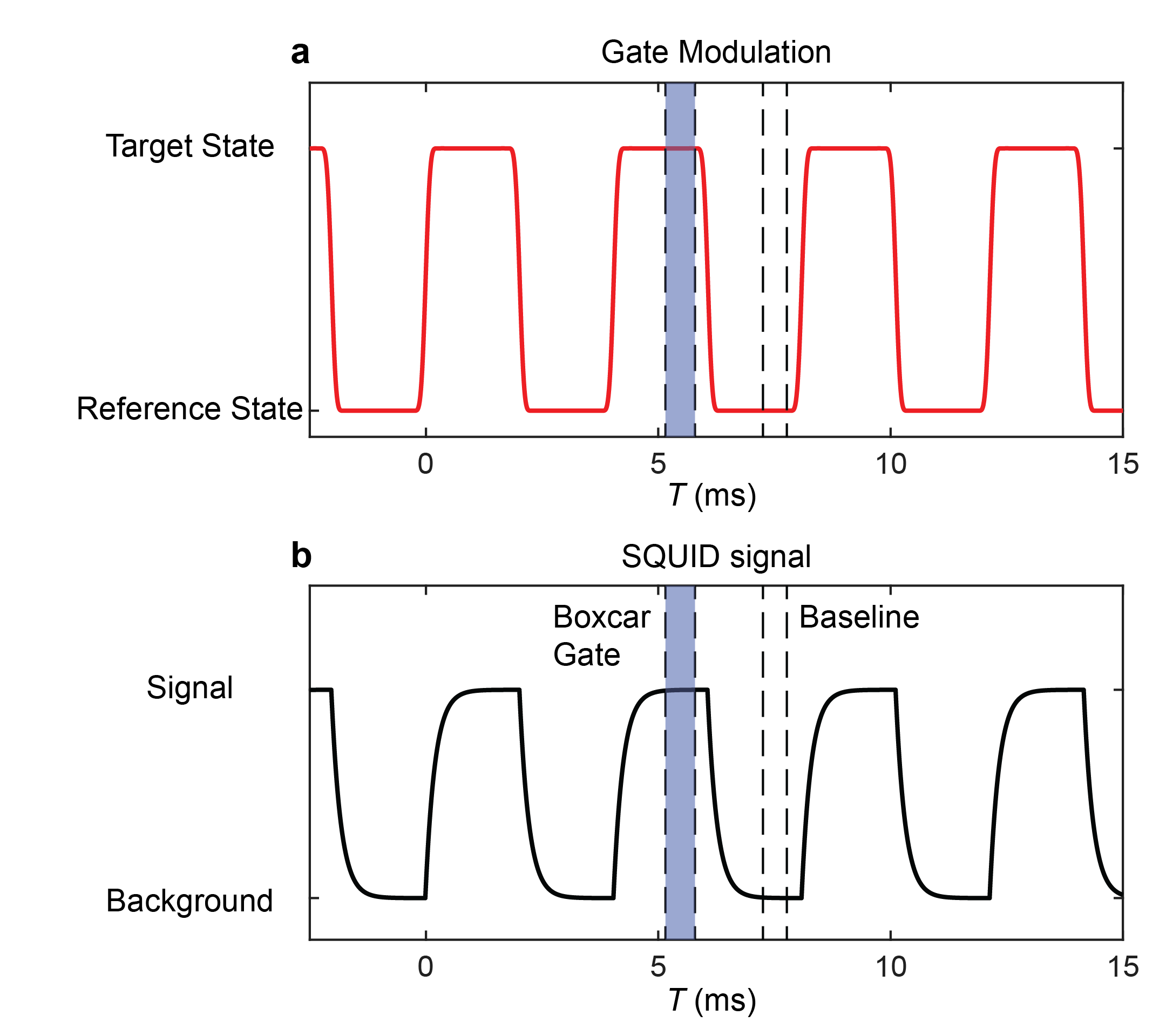}
\caption{
\textbf{Boxcar measurement.}
\textbf{a}, Illustration of square-wave gate modulation applied to the bottom gate.
\textbf{b}, Illustration of the expected nSOT signal corresponding to the gate modulation considering finite settling time effects.
}
\label{fig:Boxcar}
\end{figure*}

\begin{figure*}[ht!]
\centering
\includegraphics[width=13.7cm]{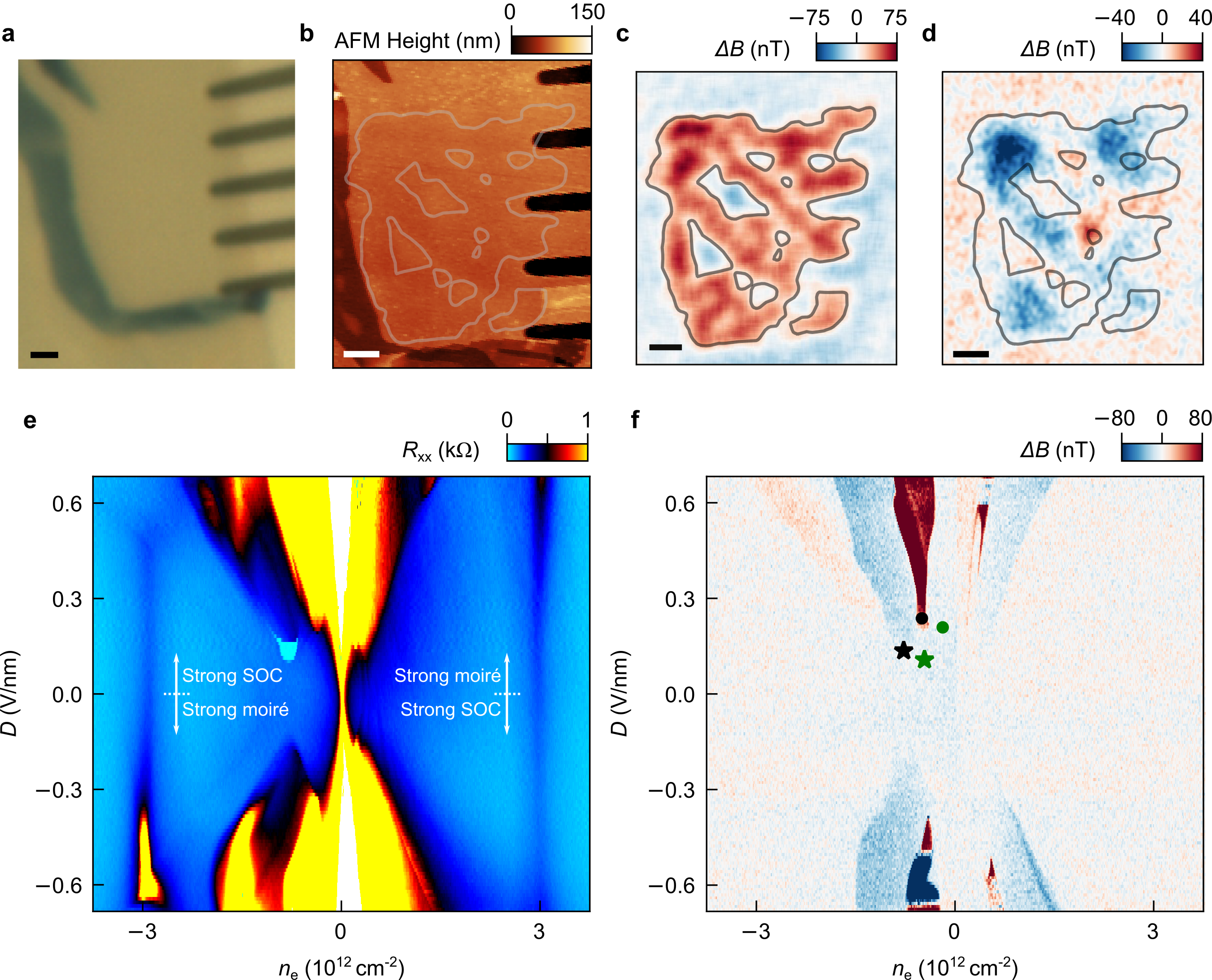}
\caption{\textbf{Device characterization and large-scale phase diagrams.} 
\textbf{a}, Optical micrograph of device, with scalebar 1 $\mu$m. 
\textbf{b}, AFM image of device, also with scalebar 1$\mu$m. Gray lines show a contour of constant $\Delta B$ measured in the valley imbalanced magnetic phase for comparison with panels c-d. Small topographical features can be observed within the otherwise smooth device region which correlate with locations of weaker ferromagnetism (panel c).
\textbf{c-d}, Data from Fig. \ref{fig:fig1}a and Fig. \ref{fig:fig2}d with the contours of constant $\Delta B$ in panel c overlaid, for comparison of vortex locations in panel d with panels b-c. 
\textbf{e}, Longitudinal resistance $R_{xx}$ vs. $n_e$ and $D$ at zero magnetic field and base temperature, with no tip-induced heating ($T\approx 30$ mK). Resistive features at high density are due to the moire pattern formed by hBN alignment \cite{patterson_superconductivity_2025}, though these do not have a significant effect on the measured fringe field (panel f).
\textbf{f}, $\Delta B$ as a function of $n_e$ and $D$ at a location in the device with sensitivity to both in-plane and out-of-plane moments (\cite{patterson_superconductivity_2025}, Extended Data Fig. \ref{fig:SuppFigSpinCanting}). Black star (dot) denotes the gate condition for superconductivity (valley imbalanced phase) images, while the green symbols are the reference points for the square wave oscillation. For this dataset, the square wave references $(n_e = 0,D)$ for each position $(n_e, D)$ in the phase diagram.
}

\label{fig:PhaseDiagrams}
\end{figure*}

\begin{figure*}[ht!]
\centering
\includegraphics[width=18cm]{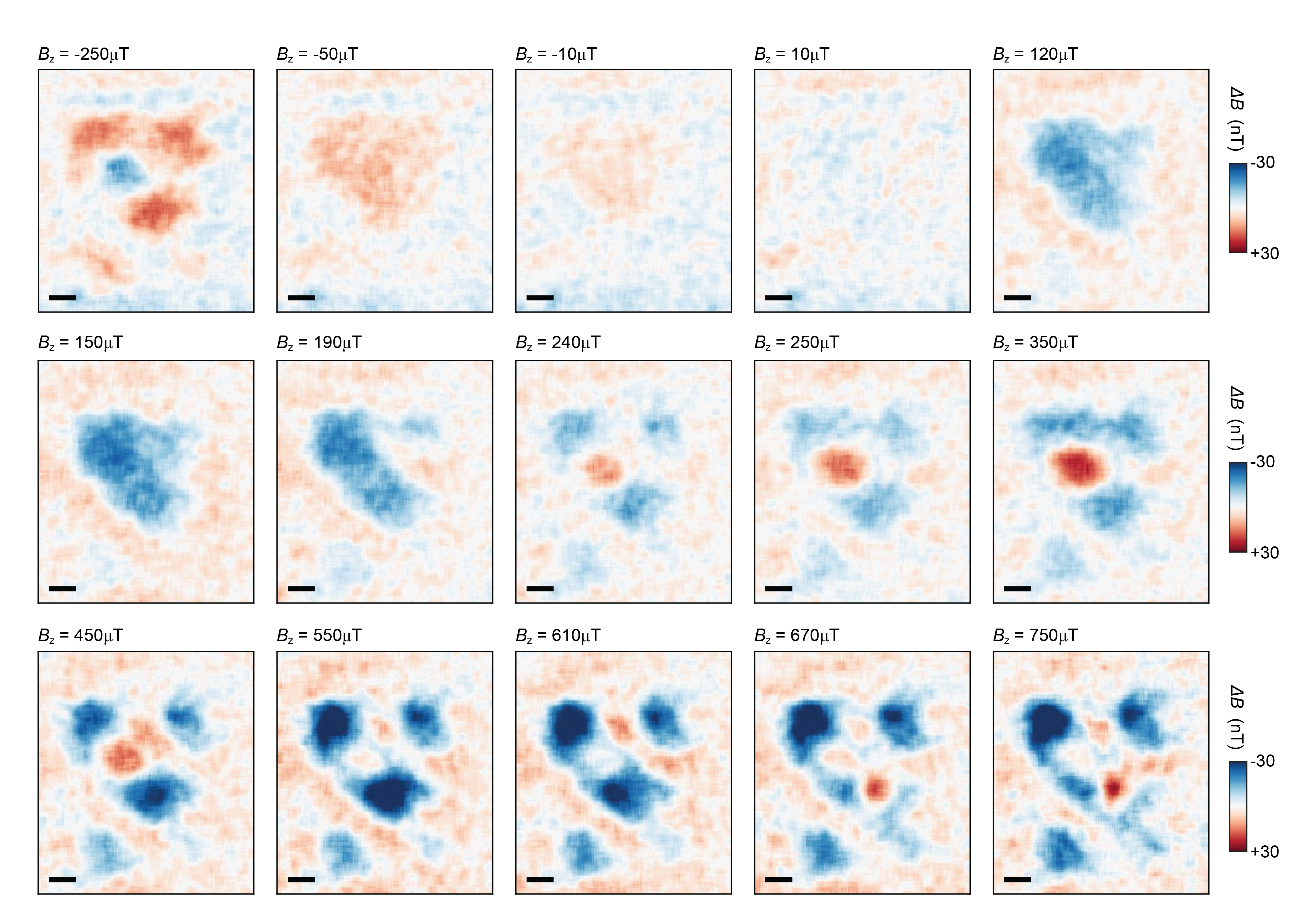}
\caption{\textbf{Spatial imaging of the Meissner effect across the entire measured $B_z$ range at $T = 210\,\mathrm{mK}$.}
}
\label{fig:all_b_dep}
\end{figure*}

\begin{figure*}[ht!]
\centering
\includegraphics[width=11cm]{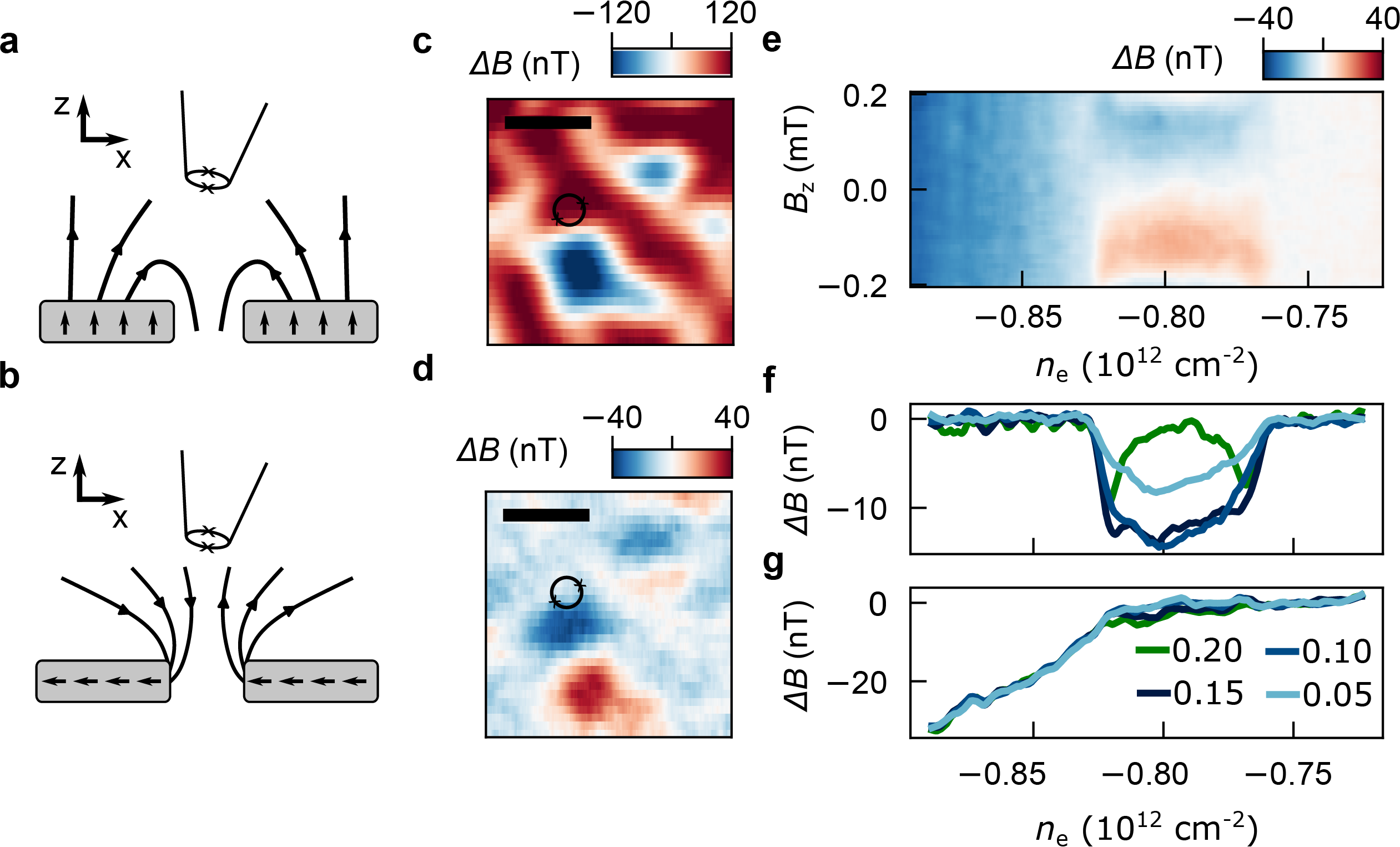} 
\caption{
\textbf{Sensitivity to in-plane spin moments and symmetrization of $\Delta B$.}
\textbf{a} Schematic of magnetic field induced by a magnetic phase with spins oriented in the $z$-direction around a magnetically `hole' region in the sample, which might be caused by a physical hole or disorder suppressing magnetic order. Field lines wrap through the `hole,' which lead to a sign reversal in the magnetic field measured by the nSOT above.
\textbf{b}, Same as panel a but for spins oriented in the plane of the sample. The sign reversal now occurs across the two edges of the `hole,' depending on the direction of spin-polarization. 
\textbf{c}, Spatial image of $\Delta B$ in the quarter-metal phase in which spin order is out of the plane.
\textbf{d}, Spatial image of $\Delta B$ in the half-metal phase in which spin order is directed in the plane. In both panels c-d, scalebars are 1 $\mu$m.
\textbf{e}, Measurement of $\Delta B$ as a function of $B_z$ at a single position (approximately marked in panels c-d by the SQUID symbol). The data  is taken along the trajectory in the $n_e$-$D$ plane noted in Fig. \ref{fig:fig_spincanting}
\textbf{f-g}, Anti-symmetrization (\textbf{f}) and symmetrization (\textbf{g}) of the data in panel e at various out of plane magnetic fields $B_z$, in mT.  
}

\label{fig:SuppFigSpinCanting}
\end{figure*}

\begin{figure*}[ht!]
\centering
\includegraphics[width=15cm]{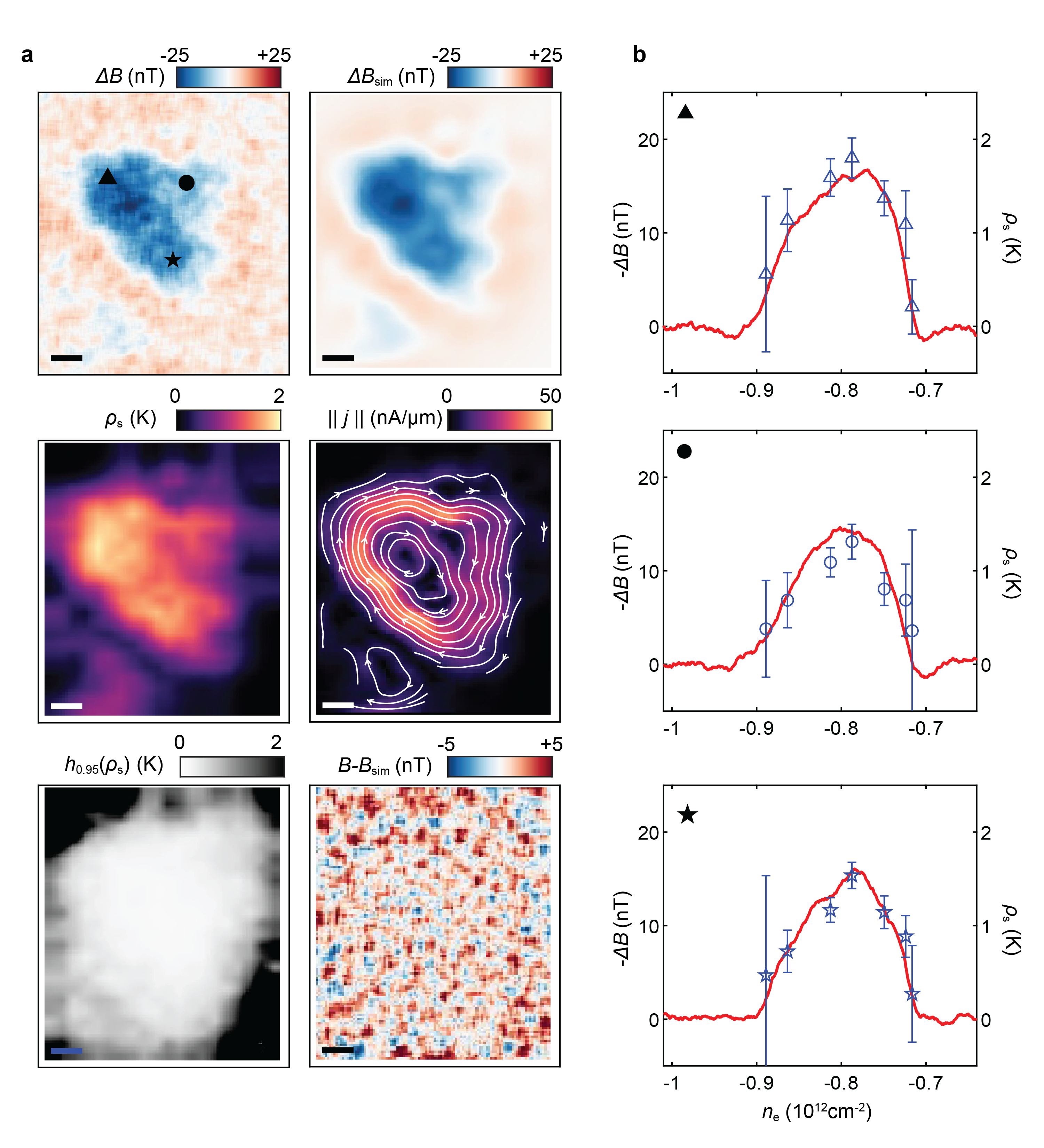}
\caption{\textbf{Additional $\rho_s$ reconstruction data}
\textbf{a}, Spatial map of $\Delta B$, $\Delta B_{\text{sim}}$, $\rho_s$, supercurrent density $||j||$, $h_{0.95}(\rho_s)$, and  fitting error $B-B_{\text{sim}}$ at $T=$210mK with displacement field $D=0.138V/nm$ and carrier density $n_e=-0.79\times10^{12}cm^{-2}$.
\textbf{b}, Correspondence between measured $\Delta B$ and reconstructed $\rho_s$ at the spatial locations marked in panel a by solid triangle, circle, and star, respectively.
}
\label{fig:all_n_dep}
\end{figure*}

\begin{figure*}[ht!]
\centering
\includegraphics[width=18cm]{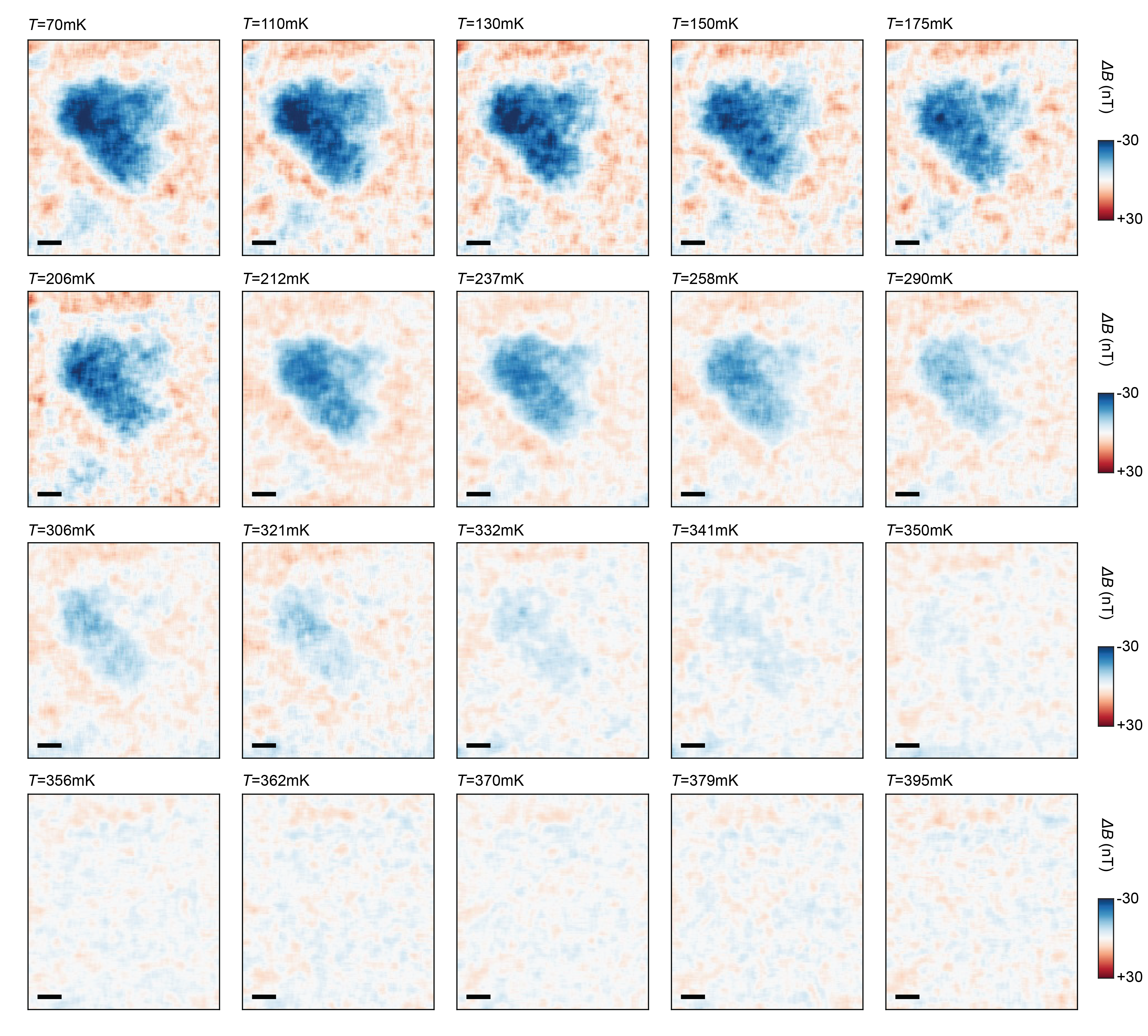}
\caption{\textbf{Spatial imaging of the Meissner effect versus temperature at $B_z = 150\mu T$.}
}
\label{fig:all_t_dep}
\end{figure*}

\begin{figure*}[ht!]
\centering
\includegraphics[width=18cm]{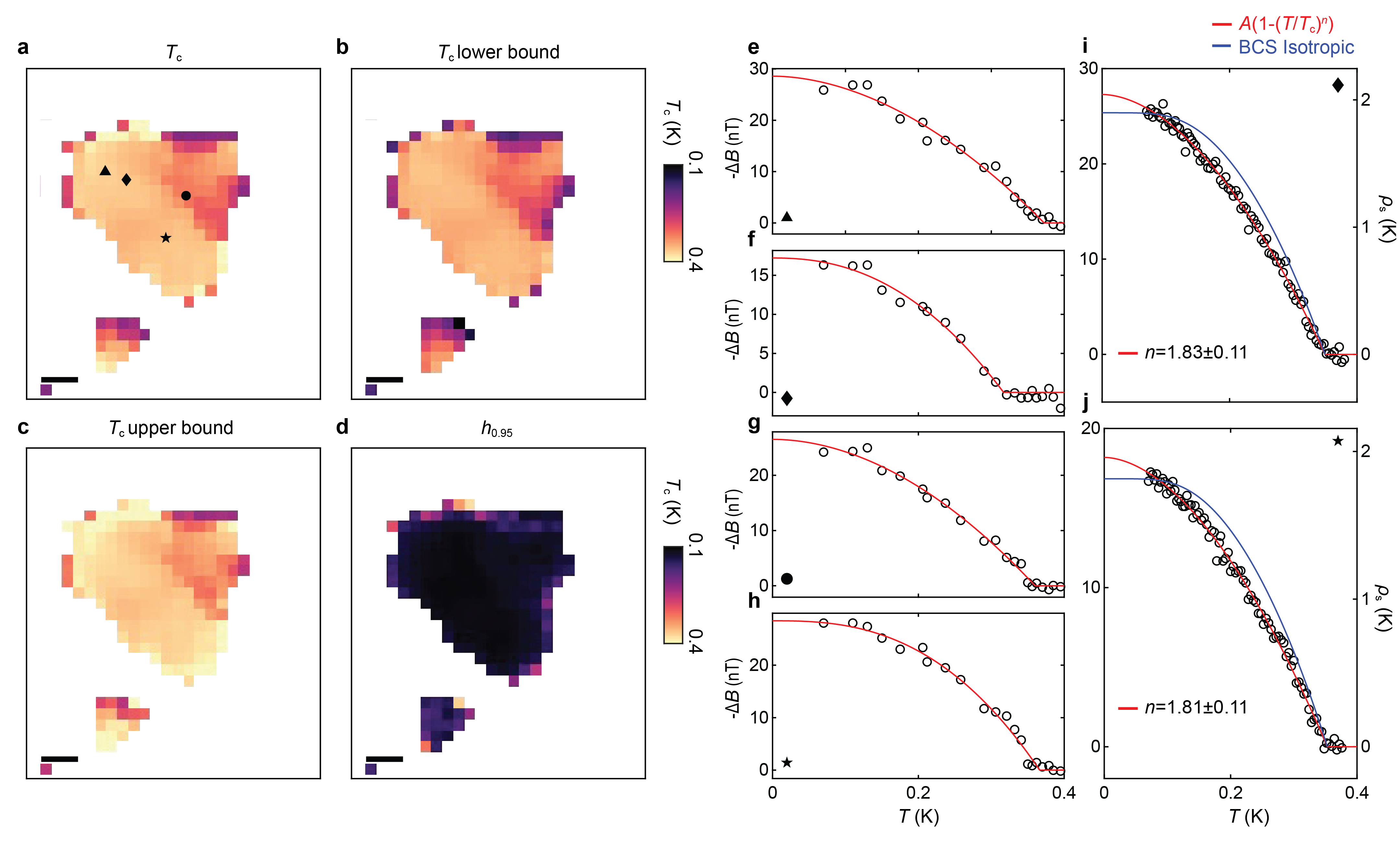}
\caption{\textbf{Additional temperature dependence data.}
\textbf{a}, Spatial map of $T_c$ obtained by fitting local $\Delta B$ with $T$.
\textbf{b}, Lower bound of the confidence interval for the fits shown in panel a.
\textbf{c}, Upper bound of the confidence interval for the fits shown in panel a.
\textbf{d}, The half width of 95\% condidence interval of the fitting in panels a-c.
\textbf{e--h}, Examples of local $T_c$ fitting at positions indicated in panel a.
\textbf{i--j}, Additional high-resolution $\Delta B(T)$ or $\rho_s(T)$ curves at the spatial locations marked by diamond and star in panel a.
}
\label{fig:tcmap_ana}
\end{figure*}

\begin{figure*}[ht!]
\centering
\includegraphics[width=18cm]{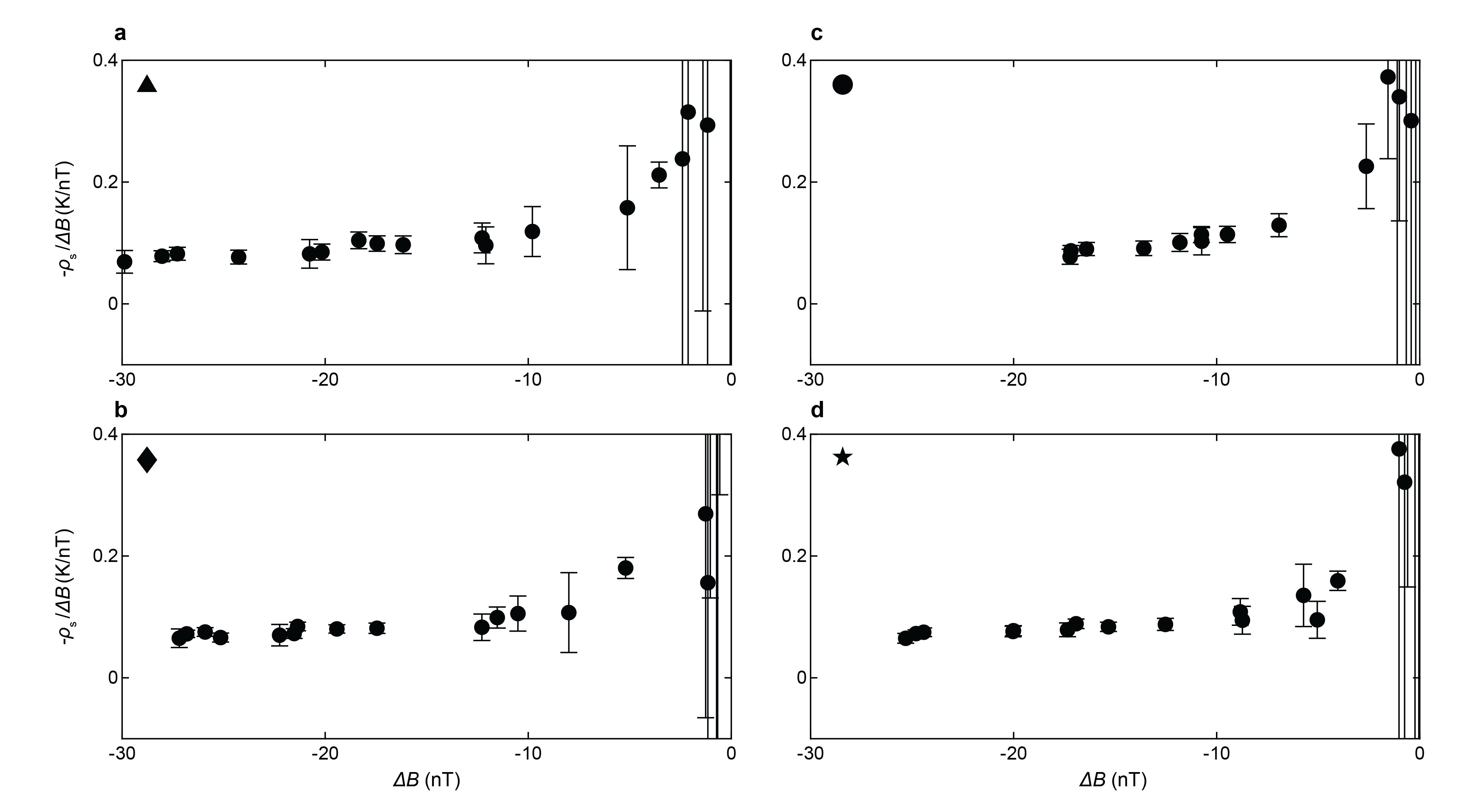}
\caption{\textbf{Linearity between $\rho_s$ and $\Delta B$ in the full temperature range.}
\textbf{a-d}, The $-\rho_s/\Delta B$ versus measured $\Delta B$ at locations labeled with solid triangle, diamond, circle, and star in Extended Data Fig. \ref{fig:tcmap_ana}a. The error bar shows the confidence interval of $\rho_s$ normalized by $-\Delta B$. 
}
\label{fig:lin_t_dep}
\end{figure*}

\begin{figure*}[ht!]
\centering
\includegraphics[width=18cm]{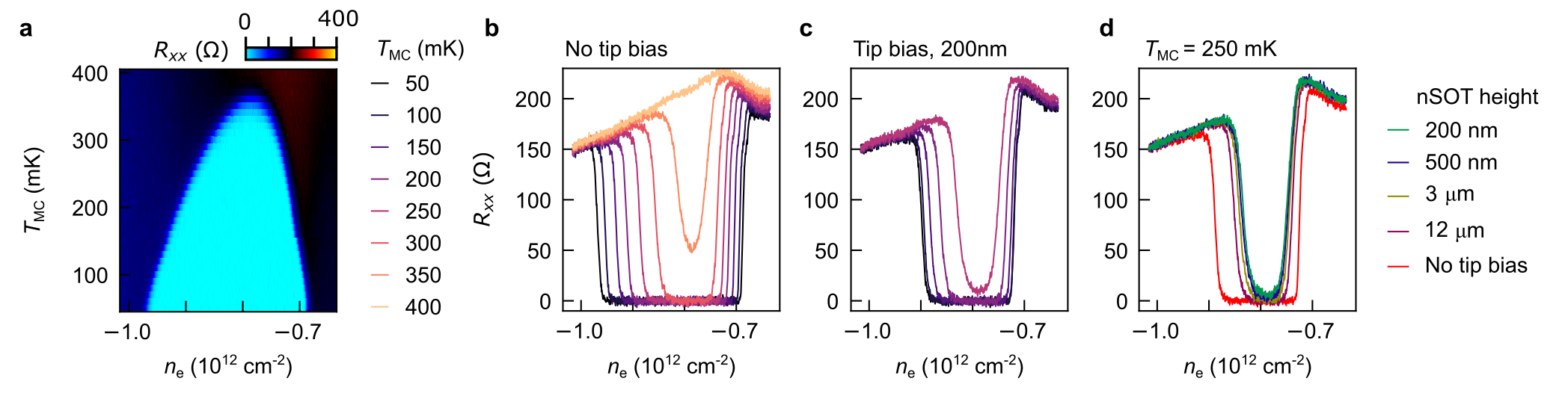}
\caption{\textbf{Transport characterization of nSOT-induced heating.} 
\textbf{(a}, $R_{xx}$ as a function of $n$ and mixing chamber temperature $T_{\rm MC}$ at $D = 0.138$ V/nm with nSOT grounded.
\textbf{b}, $R_{xx}$ as a function of $n$ at various $T_{\rm MC}$ at $D = 0.138$ V/nm with nSOT grounded.
\textbf{c}, $R_{xx}$ as a function of $n$ at various $T_{\rm MC}$ at $D = 0.138$ V/nm with the nSOT biased into the voltage state (standard measurement bias) at a height of 200 nm above the device.
\textbf{d}, $R_{xx}$ as a function of $n$ at $T_{\rm MC} = 250$ mK and $D = 0.138$ V/nm, varying the height of the nSOT biased into the voltage state. There is a weak dependence on height relative to the overall heating from the tip. 
}
\label{fig:TipHeating}
\end{figure*}

\clearpage
\renewcommand{\tablename}{\textbf{Extended Data Table}}

\begin{table*}[t]
\caption{\textbf{Summary of nSOT measurement conditions.} Measurement conditions for each nSOT measurement.
 In order, we report the measured distance between the surface of the device and the nSOT tip, the temperature of the mixing chamber $T_{MC}$, the applied $B_z$, the size of the square wave voltage $\Delta V_{\rm{bg}}$, whether the sample cell is under vacuum or in He exchange gas, and the calibrated temperature $T_{actual}$ accounting for tip heating (see Methods).}
\begin{ruledtabular}
\begin{tabular}{ccccccc}
Panel 
& nSOT-device dist. (nm) 
& $T_{\mathrm{MC}}$ (mK) 
& $B_z$ ($\mu$T) 
& $\Delta V_{\mathrm{bg}}$ (V) 
& Cell 
& $T_{\mathrm{actual}}$ (mK) \\
\hline
Fig. 1c & 150 & 50 & +150 & -0.9 & Vacuum & 70 \\
Fig. 1e & 200 & 50 & +150 & +0.35 & Vacuum & 70 \\
Fig. 1g & 200 & Variable & +150 & +0.35 & Vacuum & Variable \\
Fig. 2a,c,d,e & 200 & 50 & Variable & +0.35 & Exchange gas & 210 \\
Fig. 3a & 150 & 50 & +750 & -0.9 & Exchange gas & 210\\
Fig. 3c & 150 & 50 & $\pm$ 100 & -0.9 & Exchange gas & 210\\
Fig. 4b & 200 & 50 & +150 & +0.35 & Vacuum & 70 \\

Fig. 5a,b,d & 200 & Variable &+ 150 & +0.35 & Vacuum & Variable\\

Extended Data Fig. 3c-d & 200 & 50 & +150 & +0.35 & Vacuum & 70 \\
Extended Data Fig. 3f & 150 & 50 & +250 & Variable ($n_e = 0$) & Exchange gas & 210 \\
Extended Data Fig. 4 & 200 & 50 & Variable & +0.35 & Exchange gas & 210 \\ % all B dep
Extended Data Fig. 5c-d & 200 & 50 & +250 & -0.9 & Exchange gas & 210\\ % spin canting
Extended Data Fig. 5e & 150 & 50 & Variable & -0.9 & Exchange gas & 210\\
Extended Data Fig. 6a & 200 & 50 & +150 & +0.35 & Exchange gas & 210 \\
Extended Data Fig. 6b & 150 & 50 & +150 & +0.35 & Exchange gas & 210 \\
Extended Data Fig. 7 ($T < 210$ mK) & 200 & Variable & 150 & +0.35 & Vacuum & Variable\\ %all T dep
Extended Data Fig. 7 ($T > 210$ mK) & 200 & Variable & 150 & +0.35 & Exchange gas & Variable\\
Extended Data Fig. 8i-j & 200 & Variable & 150 & +0.35 & Vacuum & Variable\\ % const T dep

% Add more rows as needed
\end{tabular}
\end{ruledtabular}
\end{table*}

\end{document}